\documentclass{aastex6}

\newcommand{\changesbf}[1]{{#1}}

\newcommand{\orionkl}{Orion~KL}
\newcommand{\oklhc}{Orion~KL Hot~Core}

\newcommand{\iras}{IRAS~16293}
\newcommand{\irasfull}{IRAS~16293$-$2422}
\newcommand{\Xin}{$X_{\rm{in}}$}
\newcommand{\Xout}{$X_{\rm{out}}$}
\newcommand{\Tjump}{$T_{\rm{jump}}$}
\newcommand{\Nratio}{\ce{^{14}N/^{15}N}}

\usepackage[version=4]{mhchem}
\usepackage{outlines}
\usepackage{mathrsfs}

\begin{document}


\title{Exploring the Origins of Earth's Nitrogen:
Astronomical observations \\ of Nitrogen-bearing Organics in \changesbf{Protostellar} Environments}

\author{Thomas S. Rice\altaffilmark{1}}
\author{Edwin A. Bergin} 
\affil{University of Michigan Department of Astronomy \\
311 West Hall, 1085 South University Avenue,  \\
Ann Arbor, MI 48109, USA}
\altaffiltext{1}{tsrice@umich.edu}

\author{Jes K. J\o rgensen}
\affil{Centre for Star and Planet Formation,\\
Niels Bohr Institute \& Natural History Museum of Denmark, \\
University of Copenhagen, \\
\O ster Voldgade 5-7, 1350, Copenhagen K., Denmark}

\and

\author{\changesbf{S. F. Wampfler}}
\affil{Center for Space and Habitability, University of Bern, Gesellschaftsstrasse 6, CH-3012 Bern, Switzerland}

\begin{abstract}

It is not known whether the original carriers of Earth's nitrogen were molecular ices or refractory dust.
To investigate this question, we have used data and results of \textit{Herschel} observations towards two protostellar sources:
the high-mass hot core of \orionkl{}, and the low-mass protostar \irasfull{}.
Towards \orionkl{}, our analysis of the molecular inventory of \citet{Crockett14b} indicates that HCN is the organic molecule that contains by far the most nitrogen, carrying $74_{-9}^{+5}\%$ of nitrogen-in-organics.
Following this evidence, we explore HCN towards \irasfull{}, which we consider a solar analog.
Towards \irasfull{}, we have reduced and analyzed \textit{Herschel} spectra of HCN, and fit these observations against ``jump'' abundance models of \irasfull{}'s protostellar envelope.
We find an inner-envelope HCN abundance \Xin{}$ = 5.9\pm0.7 \times 10^{-8}$ and an outer-envelope HCN abundance \Xout{}$ = 1.3 \pm 0.1 \times 10^{-9}$. 
We also find the sublimation temperature of HCN to be \Tjump$ = 71 \pm 3$~K;
this measured \Tjump{} enables us to predict an HCN binding energy $E_{\textrm{B}}/k = 3840 \pm 140$~K.
Based on a comparison of the HCN/\ce{H2O} ratio in these protostars to N/\ce{H2O} ratios in comets, we find that HCN (and, by extension, other organics) in these protostars is incapable of providing the total bulk N/\ce{H2O} in comets.
We suggest that refractory dust, not molecular ices, was the bulk provider of nitrogen to comets.
However, interstellar dust is not known to have \ce{^{15}N} enrichment, while high \ce{^{15}N} enrichment is seen in both nitrogen-bearing ices and in cometary nitrogen. 
This may indicate that these \ce{^{15}N}-enriched ices were an important contributor to the nitrogen in planetesimals and likely to the Earth.

\end{abstract}



\section{Introduction}
\label{sec:intro}

Earth's nitrogen is a key element for life as well as the primary component of our atmosphere.
Because of the high volatility of the key nitrogen carriers in the interstellar medium (atomic N and \ce{N2}), it is expected that the building blocks of terrestrial worlds would be nitrogen-poor;
as a consequence, the origin of Earth's nitrogen is uncertain.
The most likely origin scenario is impact delivery by asteroids (probed by meteorites) or icy planetesimals (e.g., comets; \citealt{morbidelli12}).
The most recent work suggests an origin in delivery from chondritic meteorites, with a smaller role played by comets \citep[][and references therein]{marty16,alexander17}.

In chondrites (the most primitive meteorites), nitrogen is contained in a macromolecular organic form that is insoluble by both acids and solvents, typically known as insoluble organic matter (IOM). 
As chondrites are the most primitive and most volatile-rich of the meteorites, chondritic IOM is likely the source of the carbon, nitrogen, noble gases, and much of the hydrogen that were delivered to terrestrial planets \citep{alexander12,marty12,marty13,alexander17}.
The IOM or its chemical precursors likely formed in the outer solar system and/or the interstellar medium \citep{charnley08}, but
it is unknown whether the IOM has a primarily solar or pre-solar origin, or a mix \citep[][and references therein]{alexander17}.
\citet{bergin14,Bergin15} have discussed the astronomical origins of terrestrial carbon, arguing that the initial carriers of terrestrial carbon and nitrogen are organic ices and macromolecular organic material, including dust.

Nitrogen is contained in a variety of molecular and physical carriers in interstellar space, each of which has its own properties -- including its volatility (i.e., its sublimation temperature) and its ability to undergo isotopic fractionation (i.e., change its \ce{^{14}N/^{15}N} ratio).
Little is known about which initial carriers could have brought nitrogen into chondritic IOM, but two aspects are relevant.
First, while the most volatile forms of nitrogen (atomic N and molecular \ce{N2}) likely contain the vast majority of nitrogen atoms in the dense interstellar medium \citep[e.g.,][]{vandishoeck93,schwarz14},
their volatility prevents them from readily incorporating into solid material, so the IOM's nitrogen likely came from carriers with lower volatility.
Second, the nitrogen of the IOM is known to be highly enriched \ce{^{15}N} compared to the ISM standard \citep{furi15,alexander17}.
These factors lead us to consider two families of nitrogen carriers: nitrogen-bearing molecular ices, especially organics, and nitrogen contained in interstellar dust solids (i.e., refractory nitrogen).

Refractory nitrogen has not been directly observed. 
However, depletion in the gas-phase atomic nitrogen abundance in the diffuse ISM has been observationally constrained by \citet{knauth03} and \citet{jensen07}. 
This is generally interpreted as due to nitrogen incorporation into dust grains.
\citet{jones16a} discusses the role of nitrogen in models of interstellar carbonaceous dust.

Some nitrogen-bearing molecular ices, especially including hydrogen cyanide (HCN), exhibit a high \ce{^{15}N} enrichment in interstellar environments (e.g., \citealt{hily-blant13,wampfler14,guzman17}) and have a high abundance in the dense ISM (e.g., \citealt{Crockett14b,zernickel12,schoier02}, etc.).
HCN itself also has a rich chemical reactivity (as noted by, e.g., \citealt{noble13}).
Ammonia (\ce{NH3}) is another very abundant ice \changesbf{constituent}, but in the gas does not seem to show the same isotopic signature \citep{hily-blant13}.
Motivated by these factors, we here investigate HCN and the family of organic (C-bearing) molecular carriers of nitrogen in their possible role as the initial reservoir of nitrogen that ultimately arrives on terrestrial worlds.

A powerful way to probe the origins of nitrogen in the solar system is through astronomical observations of other forming planetary systems.
The success over the past decade of the \textit{Herschel} Space Observatory in submillimeter and far-infrared astronomy has brought a wealth of data and knowledge to the observational astrochemistry community, enabling new frontiers of molecular astrophysics.
In this study, we have drawn upon the data and results from these projects, especially the spectral surveys of \orionkl{} \citep{bergin10,Crockett14b} and \irasfull{} \citep{ceccarelli10,coutens12}.
The high-mass hot core of \orionkl{} exhibits a rich molecular spectrum.
It represents an environment in which the molecular ices available to planet-forming materials are revealed via sublimation, and hence can be studied astronomically.
The molecular abundance inventory presented by \citet{Crockett14b} of \orionkl{} enables an investigation of the relative abundances of many molecular species, especially (for our purposes) nitrogen-bearing organics.
The rich HIFI spectrum of \irasfull{}, a solar-type low-mass protostar, is also available in the Herschel Science Archive.
This spectrum can extend analyses from ground-based studies (e.g., \citealt{schoier02,caux11,wampfler14,jorgensen16}) and unlock the molecular content of \irasfull{}'s envelope, especially in its warm inner regions.

In this project, our goal is to study the distribution of nitrogen among different organic molecules, and to compare the protostellar abundance of HCN, the simplest nitrogen-bearing organic, to nitrogenic abundances in comets relative to water.
In Section~\ref{sec:data}, we introduce the observations and data used in this work, which consist of \textit{Herschel} observations towards \orionkl{} and \irasfull{}.
In Section~\ref{sec:hotcores}, we give an accounting of the nitrogen-bearing organic molecules in the \oklhc{}.
In Section~\ref{sec:iras}, we report on our analysis of \textit{Herschel} HIFI data towards \irasfull{}, and present our derivation of the HCN radial abundance distribution as inferred from a spherically symmetric model.
We discuss our results in Section~\ref{sec:discussion} in the context of cometary nitrogen abundances, and consider the available evidence pertaining to comets, meteorites, interstellar dust, and interstellar ices.
We present our conclusions in Section~\ref{sec:conclusions}.

\section{Data and Observations}
\label{sec:data}

We have analyzed the results of Herschel observations towards the \changesbf{protostellar} environments of \orionkl{} and \irasfull{} (henceforth \iras{}).
Each of these Herschel observations made use of the \changesbf{Heterodyne Instrument for the Far-Infrared (HIFI)} spectrograph, and included a spectral~survey with large ($\sim 1$ THz) bandwidth, fine resolution, and high sensitivity.
These observations are part of the HEXOS \citep[Herschel Observations of EXtra-Ordinary Sources;][]{bergin10} and CHESS \citep[Chemical HErschel Surveys of Star forming regions;][]{ceccarelli10} large programs.
Additional ground-based observations of \iras{} from \citet{wampfler14} and from the TIMASSS program \citep{caux11} complement the HIFI data in our analysis.

\smallskip
\noindent \textbf{\orionkl{}} --- 

We use observations of \orionkl{} from the HEXOS program \citep{bergin10}; these observations and their reduction were described in detail in \citet{Crockett14b}.
The dataset consists of a $\sim$1.2~THz-wide spectrum from 480 to 1907~GHz at a resolution of 1.1~MHz.
These data were previously presented and analyzed in \citet{crockett10,wang11,plume12,neill13b,neill13a}, and \citet{Crockett2014a,Crockett14b,Crockett15}.
In this paper, we primarily analyze the molecular abundances of the \oklhc{} presented in \citet{Crockett14b} and \citet{neill13a}; we additionally use results from the 1.3~cm survey of \orionkl{} by \citet{gong15}.

\smallskip
\noindent \textbf{\irasfull{}} --- 

We have freshly reduced and analyzed the HIFI observations of \iras{} from the CHESS program \citep{ceccarelli10}.
This spectrum spans 0.9~THz from 480--1800~GHz, with gaps, and again has a spectral resolution of 1.1~MHz.
These data were previously presented and analyzed in \citet{hily-blant10,bacmann10}, and \citet{coutens12}.
We obtained the \textit{Herschel} HIFI spectra of \iras{} from the Herschel Science Archive (\url{http://archives.esac.esa.int/hsa/whsa/}) and downloaded the Level 2.5 products. 
We averaged the horizontal and vertical polarizations in GILDAS/CLASS to improve the signal-to-noise.
Finally, we resampled all lines to a common velocity resolution of 0.7 km~s$^{-1}$, the resolution of the lowest-$J$ line, in order to achieve consistency between lines and improve the per-channel signal-to-noise.

We have complemented our HIFI data of \iras{} with two sets of ground-based data covering lower-energy transitions.
We have used observations of \iras{} at 260 and 345~GHz from the Atacama Pathfinder EXperiment (APEX) 12-meter telescope in Chile.
\changesbf{These APEX data were previously presented} in \citet{wampfler14}, where a detailed description of these data appears.
Finally, for the 86~GHz $J=1-0$ transition of \ce{H^{13}CN}, we have used the publicly available reduced TIMASSS data \citep{caux11}.
\section{Nitrogen-bearing molecules in the Orion KL Hot Core}
\label{sec:hotcores}

\subsection{Background}

\citet{Crockett14b} analyzed the HEXOS spectra of \orionkl{} and identified $\sim13,000$ lines coming from 39 molecules (79 distinct isotopologues).
This study made use of IRAM-30m data (80-280~GHz) to constrain low-energy molecular transitions, 
and of an ALMA interferometric spectral survey at 214-247~GHz to constrain both the millimeter spectrum and the spatial structure of the continuum and molecular emission. 
Spatial information is especially important due to the multiple physical components within \orionkl{} which are not spatially resolved by \textit{Herschel}.
\changesbf{\citeauthor{Crockett14b} present ALMA maps of various molecules, indicating that, e.g., \ce{CH3CN} and \ce{CH3OH}, fill the Hot Core's spatial extent as traced by its continuum emission.
}

\citeauthor{Crockett14b} modeled the emission from each component individually.
\changesbf{Temperature gradients within the Hot Core were approximated using the following approach:
A single 10\arcsec{} component was fit to the emission of each molecule. 
If a single-temperature component was not sufficient to reproduce the observed emission, then additional subcomponents were added, either larger and cooler or smaller and hotter, in size steps of factor two.}
\citeauthor{Crockett14b} use two modeling codes: a local thermodynamic equilibrium (LTE) code called XCLASS, and a non-LTE, large velocity gradient (LVG) code called MADEX \citep{cernicharo12}.
These emission modeling codes were used to derive column densities and abundances for all observed species.
The water abundances, in particular, were derived by \citet{neill13a}.
Among the physical components of \orionkl{}, we have focused on the Hot Core as the most relevant astrochemical environment in which ices are evaporating, revealing the total molecular budget available at this stage of star and planet formation; in the other regions, many molecules which are important to planetesimal formation are hidden in ices.

\changesbf{Additional abundances of nitrogen-bearing organics in \orionkl{} are available in the recent literature. 
Formamide (\ce{NH2CHO}) was measured by \citet{adande13} towards \orionkl{} with an abundance of $5 \times 10^{-11}$, but because the Hot Core was not spatially or kinematically resolved in their measurement, we omit formamide from our analysis.
\citet{Crockett14b} identified \ce{NH2CHO} emission only towards the compact ridge of \orionkl{}, and not the Hot Core, further justifying this exclusion.
In any case, its abundance appears to be far below that of HCN.
Another N-bearing organic, methyl isocyanate (\ce{CH3NCO}), was measured in \orionkl{} by \citet{cernicharo16}, who identified abundances towards distinct spatial positions within \orionkl{}.
They measured the \ce{CH3NCO} abundance (relative to \ce{H2O}) towards a position colocated with the Hot Core. 
By scaling the fractional abundance of \ce{CH3NCO} relative to HCN and \ce{CH3CN} found by \citet{cernicharo16} to the abundances of HCN and \ce{CH3CN} in the \oklhc{} found by \citet{Crockett14b}, we estimate the \ce{CH3NCO} abundance relative to \ce{H2} in the Hot Core to be in the range $3 \times 10^{-9} - 1 \times 10^{-8}$.}

\changesbf{
The cyanide radical (CN) is a common molecule in the cool and relatively less-dense molecular ISM, but is not abundant in the \oklhc{}.
CN is destroyed rapidly in hot cores via reactions with \ce{H2}, oxygen (O and \ce{O2}), and simple hydrocarbons such as \ce{C2H2} and \ce{C2H4} \citep{tielens13,whittet2013planetary}.
Accordingly, CN is not detected by \citet{Crockett14b} in either the Hot~Core or the compact ridge; they find it only towards the plateau and extended ridge regions of \orionkl{}, where its low rotation temperature of $\sim 20-40$~K indicates that CN emission probes cool material.
}

Ammonia (\ce{NH3}) is another nitrogen carrier of interest.
\citet{Crockett14b} were unable to derive an abundance of \ce{NH3} towards \orionkl{} as the \textit{Herschel} survey did not detect lines spanning a sufficiently large upper-state energy coverage to constrain the abundance and excitation state of the gas.
\ce{NH3} was studied observationally towards \orionkl{} by \citet{gong15}, who performed a 1.3~cm line survey of the region with the Effelsberg-100~m telescope; their results are noted below alongside those of \citet{Crockett14b}.

\subsection{Analysis}

\begin{figure*}
\plottwo{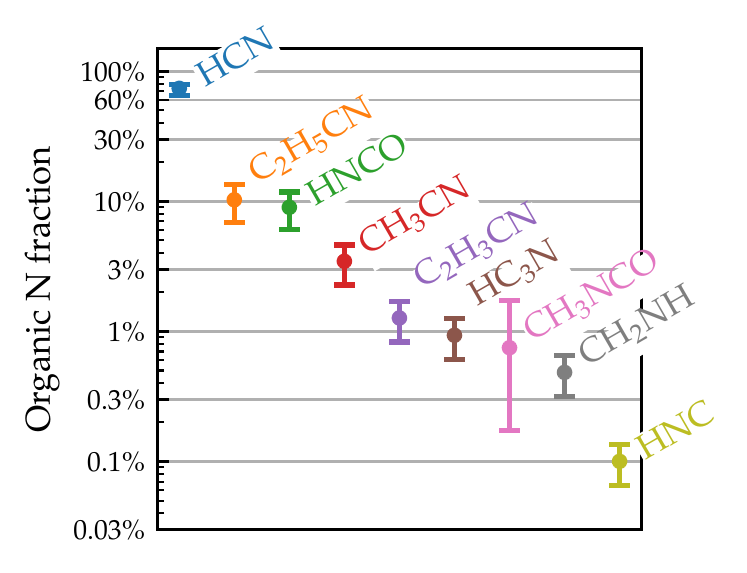}{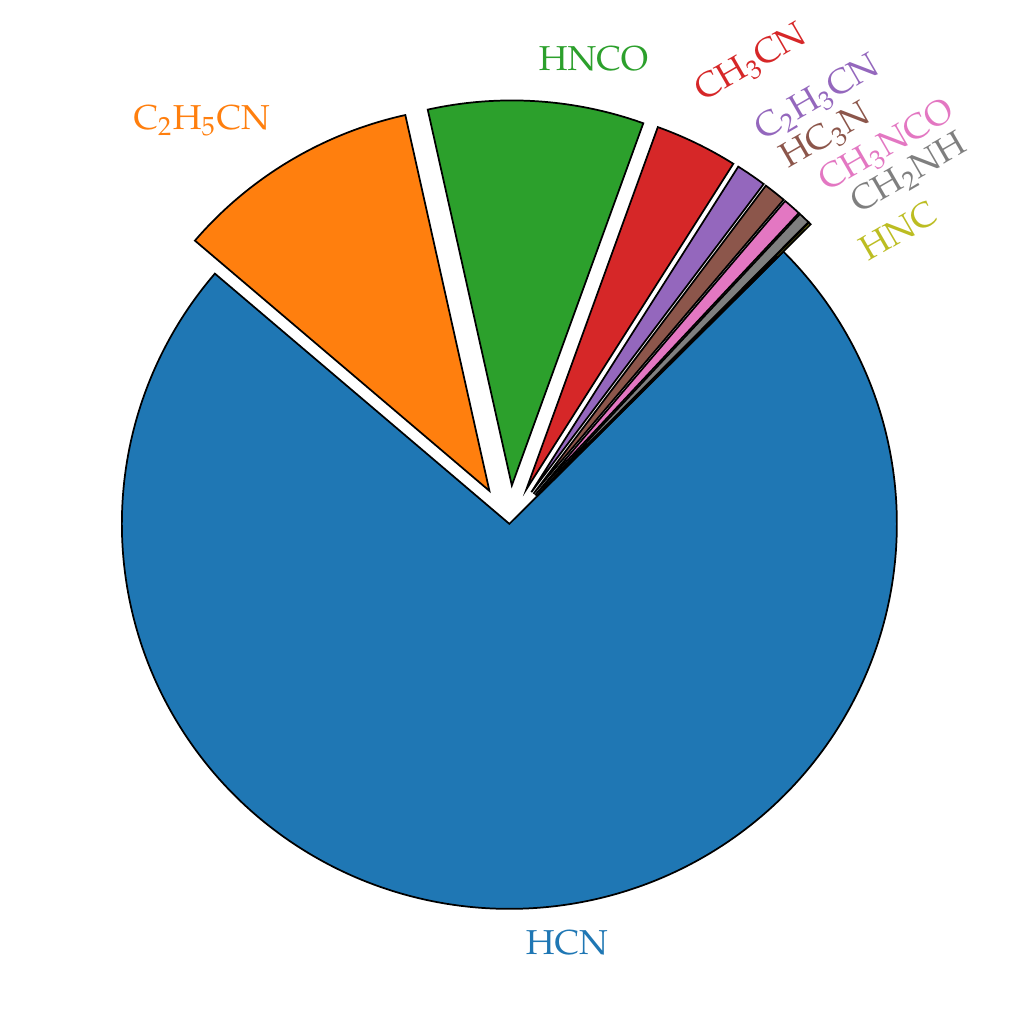}
\caption{Accounting of organic nitrogen-bearing molecules in Orion KL Hot Core, from abundances presented in \citet{Crockett14b} \changesbf{and (for \ce{CH3NCO}) \citet{cernicharo16}}. Left: Fraction of total nitrogen found in organic species, per molecule. Right: Pie chart of the mean fraction for each species. \changesbf{The larger errors for \ce{CH3NCO} are described in the text.} \label{fig:new_nitrogen_fraction_OKL}}
\end{figure*}

In order to study the composition of nitrogen-bearing organic (C-containing) molecules,
we have analyzed the molecular abundances presented by \citet{Crockett14b}.
To do this, we isolated the abundances ($N(\ce{X})/N_{\ce{H2}}$, where $N$ represents a column density in cm$^{-2}$) of all compounds containing both carbon and nitrogen from Table~8 of \citet{Crockett14b}.
We calculate the total amount of organic nitrogen by adding all N atoms contained in organic molecules\footnote{In principle, molecules with multiple nitrogen atoms would count multiply to this total, in proportion to the number of N atoms they possessed; however, all N-bearing organics in this table have just one N atom, so this sum is identical to the amount of N-bearing organic molecules.}.
Each molecule's fraction of organic nitrogen corresponds to the number of N atoms contained in that molecule, divided by the sum of N in organics.
\citet{Crockett14b} estimate that individual column density measurements have an uncertainty of 25\%.
We estimate uncertainties on the relative abundances to be 35\%, by taking each molecule's column density uncertainty to be independent and adding their uncertainties in quadrature.
\changesbf{For \ce{CH3NCO}, we assign an additional $\sim$50\% uncertainty on the higher and lower ends of the abundance range scaled from the measurements in \citet{cernicharo16}, by adding in quadrature the relative errors on column densities in their measurements to those in \citet{Crockett14b}; this accounts for the differences in observing setups.}
We show the distribution of N atoms among organic species in Figure~\ref{fig:new_nitrogen_fraction_OKL}, on a logarithmic percentage scale with appropriate error bars (left) and in a linear pie-chart form (right).

\subsection{Results}

The total abundance (relative to \ce{H2}) of N atoms in organic (C-containing) molecules in \oklhc{} is $8.6 \pm 3.4 \times 10^{-7}.$
HCN, with an abundance of $X(\ce{HCN}) = 6.4 \pm 2.5 \times 10^{-7}$, is the most abundant N-bearing organic, containing $74 ^{+5}_{-9}\%$ of all N atoms that are in organic molecules.
The next most abundant species are \ce{C2H5CN} (ethyl cyanide), with abundance $8.9 \pm 3.5 \times 10^{-8}$, containing $10 \pm 3 \%$ of N atoms within organic molecules; HNCO (isocyanic acid), with abundance $7.8 \pm 3.1 \times 10^{-8}$, containing $9 \pm 3 \%$ of organic N atoms; and \ce{CH3CN} (methyl cyanide), with abundance $3.0 \pm 1.2 \times 10^{-8}$, containing $3 \pm 1 \%$ of organic N atoms. 
The remaining observed species (\ce{C2H3CN}, \ce{HC3N}, \changesbf{\ce{CH3NCO},} \ce{CH2NH}, and \ce{HNC}) each \changesbf{constitute} $\lesssim 1\%$ of the nitrogen budget among organic molecules.
Therefore, HCN accounts for more than two-thirds of the organic N budget.

Some abundant N-bearing non-organic species are excluded from the above analysis but have known abundances.
As noted above, \ce{NH3} was observed by \citet{gong15}.
Their measurements of many \ce{NH3}, \ce{^{15}NH3}, and \ce{NH2D} lines suggest an \ce{NH3} abundance in the range $0.8 - 6 \times 10^{-6}$.
This range indicates that the \ce{NH3} abundance is at least equal to, and possibly an order of magnitude higher than, the abundance of HCN in the \oklhc{}.
The NO abundance derived by \citet{Crockett14b} is $5.5 \times 10^{-7}$, which is consistent (within the errors) with the HCN abundance.
Also, \citet{Crockett14b} derive an NS abundance of $6.8 \times 10^{-9}$.

\subsection{Limitations, caveats, and comments of hot core analysis}

The abundance measurements of \citet{Crockett14b} depend on the accuracy of the \ce{H2} column density measured in \citet{plume12}, which was itself scaled from an empirical measurement of \ce{C^{18}O} column density using an \ce{H2}:\ce{C^{18}O} ratio of $5 \times 10^6$; therefore, any issues in CO:\ce{H2} scaling would complicate the abundance measurements.
The molecular column densities derived for the \oklhc{} observed by \textit{Herschel}, as presented in \citet{Crockett14b}, depend on assumed source sizes.
For \orionkl{}, \citet{Crockett14b} and \citet{plume12} assume the same source size for the Hot Core, so any errors should cancel out.

One issue that may complicate the interpretation of the \textit{Herschel} molecular abundance inventories is the following ambiguity.
As noted in \citet{Crockett15}, there are two ways to account for the increased abundances of many molecules in the \oklhc{}:
\begin{enumerate}
	\item Grain surface chemistry at cold temperatures that produces icy molecules which are then sublimated (and thus revealed) by the high-temperature environment of the hot core; or
	\item Hot gas chemistry that produces gas molecules in-situ, which only operates in the hot environment.
\end{enumerate}
Our interpretation assumes that the observed molecular populations are dominated by molecules which are formed under the cold conditions of option (1) and are simply \textit{revealed} by the hot environment, as opposed to molecules which \textit{form} due to the hot environment.
\citet{Crockett15} suggested two ways to probe this further.
\changesbf{First, to look for spatial gradients in the D/H ratios of complex N-bearing species, as this would be a sign of grain surface chemistry. 
Second: 
in the scenario in which the N-bearing organics are desorbing at higher temperatures than O-bearing organics, rather than forming in the hot gas itself, we would expect to see that the emission from N-bearing species is clumpier and has a higher excitation temperature on average.}

Among nitrogen-bearing organics, \citet{Crockett14b} modeled \ce{CH3CN}, HCN, and HNC with both the XCLASS (simple LTE) and MADEX (non-LTE, large velocity gradient) molecular line spectra modeling codes.
The abundances derived for \ce{CH3CN} are identical using both methods; compared to XCLASS, the HCN abundance from MADEX is a factor $1.7 \times$ higher; and the HNC abundance from MADEX is a factor $6.3 \times$ \changesbf{lower}.
The MADEX modeling is thought to yield more accurate abundances, as it is more sophisticated and can account for both temperature and density gradients, so we preferentially show MADEX results where possible. 
We note that our results do not change substantially when only XCLASS abundances are used.

Regardless of the above caveats and limitations on the absolute abundance measurements, our results clearly indicate that in a relative comparison, HCN is the dominant N-bearing organic molecule in the \oklhc{}.

\section{The HCN abundance of \irasfull{}}
\label{sec:iras}

While our analysis of \orionkl{} is valuable especially due to its rich available spectrum, \orionkl{} has a higher mass than the region in which the Sun formed, and there may be chemical differences between a high-mass hot core and the birth environment of our Sun.
To attempt to control for this difference, we have also studied the low-mass protostar \iras{}.
\iras{} is a low-mass protostar benchmark, and is among the brightest low-mass protostellar sources in molecular line emission (see \citealt{jorgensen16} for an extended review of the literature on this source).
Motivated by the high abundance of HCN among organic N-bearing molecules towards \orionkl{},
we have measured the HCN content towards \iras{}.
In this measurement, we have made use of the ground-based measurements of low-$J$ states presented in \citet{caux11} and \citet{wampfler14}, and space-based high-$J$ states from \textit{Herschel} \citep{ceccarelli10}. 
We have fit a spherically symmetric ``jump abundance'' model to the observed HCN lines, in order to derive the HCN abundance within and outside the warm sublimation zone.

\medskip
\subsection{Background}

\iras{} is a prototypical Class 0 protostellar system which has been well-characterized physically and chemically.
It lies at a distance of $\sim$ 120 pc \citep{loinard08} and has a luminosity of $21\pm5$ L$_\odot$ \citep{jorgensen16}. 
It is a protobinary, with two major components (\textit{IRAS16293A} and \textit{IRAS16293B}) that have a separation of $5.1\arcsec$ or $\sim 620$ AU \citep{looney00,jorgensen16}. 
IRAS16293A is itself a possible tight binary with separation $1\arcsec$ \citep{chandler05}.
\iras{} is notable for its rich molecular line spectrum \citep{vandishoeck95,cazaux03} and has been the target of numerous single-dish and interferometric studies, most recently the ALMA-PILS survey \citep{jorgensen16,lykke17,ligterink17}.	
Combined interferometric and single-dish imaging of HCN and \ce{HC^{15}N} lines was presented by \citet{takakuwa07}.
Within this system, an inner ``hot corino'' of $\gtrsim 100$~K gas enriched with evaporated ices is thought to be present, in order to explain molecular abundance jumps and rich organic chemistry \citep{schoier02,cazaux03}.
Consequently, it is an excellent low-mass protostellar astrochemical benchmark, especially for our interest in studying icy nitrogen materials provided to early planetesimals.

\changesbf{The nitrogen-bearing organics in the hot corino of \iras{} are not as well-constrained as those in \orionkl{}, so it is difficult to quantify whether nitrogen has major chemical differences in \iras{} versus \orionkl{}.
Nonetheless, unpublished and recently-published results on column densities of nitriles
} 
(\ce{-C#N}-bearing organics) 
\changesbf{
are consistent with HCN containing the majority of N in organics in \iras{}. 
Using the PILS data, \citet{calcutt18a,calcutt18b} studied complex nitriles in \iras{} including \ce{CH3CN}, \ce{C2H5CN}, \ce{C2H3CN}, \ce{HC3N}, and \ce{CH3NC}.
Of these, \ce{CH3CN} is by far the most abundant; it is an order of magnitude more abundant than the next most abundant species, \ce{C2H5CN}.
PILS measurements of HCN and its isotopologues are still unpublished;
an analysis of the optically-thick (on 0\farcs 5 scales) \ce{HC^{15}N} $4-3$ line at 344.2~GHz indicates that \ce{HC^{15}N} is at least $1.6 \times$ as abundant as \ce{CH3C^{15}N}, and therefore (assuming that \ce{^{15}N} fractionation is comparable between HCN and \ce{CH3CN}) that HCN is at least $1.6 \times$ as abundant as \ce{CH3CN}, and therefore the dominant organic N carrier in \iras{}.
}


\floattable
\begin{deluxetable}{cDDccc}
\tablecaption{Lines of \ce{H^{13}CN} used to fit model. \label{tab:model_lines}}
\tablehead{
	\colhead{Line} & 
	\twocolhead{$E_{\textrm{up}}$} & 
	\twocolhead{Freq} & 
	\colhead{Telescope} & 
	\colhead{Pointing} &
	\colhead{Beamsize} \\
	\colhead{ } & 
	\twocolhead{(K)} & 
	\twocolhead{(GHz)} & 
	\colhead{} & 
	\colhead{offset\tablenotemark{a} (\arcsec)} &
	\colhead{HPBW (\arcsec)}
}
\decimals
\startdata
$1 - 0$ & 4.1 & 86.340 & IRAM & 5.0 & 29.1 \\
$3 - 2$ & 24.9 & 259.012 & APEX & 0.0 & 24.3 \\
$4 - 3$ & 41.4 & 345.340 & APEX & 0.0 & 18.2 \\
$6 - 5$ & 87.0 & 517.970 & HIFI & 2.5 & 41.6 \\
$7 - 6$ & 116.0 & 604.268 & HIFI & 2.5 & 35.7 \\
$8 - 7$ & 149.2 & 690.552 & HIFI & 2.5 & 31.2 \\
$9 - 8$ & 186.4 & 776.820 & HIFI & 2.5 & 27.7 \\
$10 - 9$ & 227.9 & 863.071 & HIFI & 2.5 & 25.0 \\
$11 - 10$ & 273.4 & 949.301 & HIFI & 2.5 & 22.7 \\
\enddata
\tablenotetext{a}{\changesbf{Pointing offsets from source A are described in \citet{caux11} for the IRAM data, \citet{wampfler14} for the APEX data, and \citet{coutens12} for the HIFI data.}}
\tablecomments{Gaussian fits to these lines, and their HCN and \ce{HC^{15}N} counterparts, appear in Table~\ref{tab:hcn_lines} in Appendix~\ref{sec:appendix_A}.}
\end{deluxetable}

\subsection{Herschel Data Reduction}

We fit lines of HCN, \ce{H^{13}CN} and \ce{HC^{15}N} from $J=6-5$ to $10-9$ with Gaussian profiles\footnote{We note that the \ce{HC^{15}N} $7-6$ line is blended with a stronger adjacent \ce{SO} $J=14-13$ line at 602.292 GHz; we fit the two lines simultaneously.}.
Because some of the \ce{H^{13}CN} and \ce{HC^{15}N} lines are quite weak, we have used the line widths and velocity centers of the corresponding HCN lines for either (a) an initial guess for the Gaussian fit, or (b) a fixed parameter such that only the line intensity was allowed to float in the fit, depending on the strength of the lower-intensity line.
Results of this fitting are presented in Appendix~\ref{sec:appendix_A};
Figure~\ref{fig:hcn_hifi_lines} shows the lines and their Gaussian fits, 
and the fitted line parameters are presented in Table \ref{tab:hcn_lines}.

We have estimated the optical depths from the flux ratios between the standard isotopologue HCN and its rarer isotopologue \ce{H^{13}CN}.
By assuming an isotopic ratio between HCN and \ce{H^{13}CN}, we can relate the optical depth of a given rarer isotopologue in the following way (as described by \citealt{Crockett2014a}):
\begin{equation}
	\tau_{\rm{iso}} = -\ln \left( 1- \frac{T_{\rm{iso}}}{T_{\rm{main}}} \right)
	\label{eq:tau_iso}
\end{equation}
and 
\begin{equation}
	\tau_{\rm{main}} = \tau_{\rm{iso}} \cdot \left( \frac{\ce{^{12}C}} {\ce{^{13}C}} \right).
	\label{eq:tau_main}	
\end{equation}
In these equations, $T_{\rm{main}}$ and $T_{\rm{iso}}$ denote the peak brightness temperatures of the main and rarer isotopologue lines, respectively, and $\tau_{\rm{main}}$ and $\tau_{\rm{iso}}$ likewise denote those lines' optical depths.
This equation assumes that $\tau_{\rm{main}} \gg \tau_{\rm{iso}}$, which is appropriate for comparing \ce{^{12}C} with \ce{^{13}C}.
We adopt a carbon isotopic ratio of $\ce{^{12}C/^{13}C} = 69 \pm 6$ in the local ISM \citep{wilson99}.
The optical depths $\tau$ are given in Table~\ref{tab:hcn_derived_props}.
We found optical depths $\tau_{\ce{H^{13}CN}} = 0.08 - 0.24$ (corresponding to $\tau_{\ce{HCN}} = 5.9 - 16.5$). 
Therefore, we can confirm that none of the \ce{H^{13}CN} lines is optically thick.

\subsection{Nitrogen Isotopic Fractionation}

Nitrogen isotopic fractionation seen in a low-mass protostellar system such as \iras{} can be compared to the nitrogen fractionation measured in solar system objects such as meteorites and comets \citep{hily-blant13,furi15}.
By assuming that \ce{H^{13}CN} and \ce{HC^{15}N} have the same excitation properties in \iras{}, and with knowledge (from the HCN lines) that the observed \ce{H^{13}CN} and \ce{HC^{15}N} lines are all optically thin ($\tau \lesssim 0.25$), we can calculate the \ce{^{14}N/^{15}N} ratio for each rotational state of HCN independently.
In the \textit{Herschel} data, the following lines have definite detections for both \ce{H^{13}CN} and \ce{HC^{15}N}: $6-5$, $7-6$, $8-7$, and $9-8$.
We find the \ce{^{14}N/^{15}N} ratio in each line to be $119 \pm 13$, $155 \pm 43$, $151 \pm 89$, and $210 \pm 70$ respectively.
In all cases, these are \ce{^{15}N}-enriched relative to the local ISM \ce{^{14}N/^{15}N} ratio of $388\pm38$ and the solar standard of 270 \citep{wilson99}.
These results are broadly consistent with the \ce{^{14}N/^{15}N} ratios measured by \citet{wampfler14} for this source, in which the $4-3$ and $3-2$ lines had $190 \pm 38$ and $163 \pm 20$, respectively.
The weighted mean of the \ce{^{14}N/^{15}N} values is $140\pm10$.

\subsection{Model}

In order to derive the HCN abundance profile around \iras{}, we have used spherically symmetric models of the temperature, density, and molecular abundance in order to produce model spectra and compare them with the observed emission from \ce{H^{13}CN}.
We focus on this isotopologue, which has a much lower abundance than HCN and therefore fewer issues with optical depth.
Aspects of this model fitting are similar to the approach carried out by \citet{coutens12}, who use the physical model of \iras{} derived in \citet{crimier10}.
We choose this modeling approach to derive the HCN abundance in a way that is consistent with the previous \ce{H2O} abundance modeling performed by \citet{coutens12}. 
\changesbf{This facilitates} direct comparison of these results and yield HCN/\ce{H2O} ratios,
\changesbf{though as noted by \citealt{visser13} and further discussed in Section~\ref{sec:bulk_nitrogen}, the \citeauthor{coutens12} water abundance of $5 \times 10^{-6}$ is possibly only a lower limit, with theoretical and observational support for a value closer to $10^{-4}$.}

We use a model which has two regions of constant HCN abundance: an outer ``cold'' region with a low gas-phase HCN abundance (because most HCN is frozen onto dust grains), and an inner ``warm'' region with a high gas-phase HCN abundance due to sublimation of HCN from those grains.
This ``jump abundance'' model has three free parameters:
The inner abundance ($X_{\rm{in}}$), the outer abundance ($X_{\rm{out}}$), and the temperature at which sublimation rapidly occurs ($T_{\rm{jump}}$).
Jump abundance models have been adopted to interpret observations of protostars in many papers (e.g., \citealt{schoier02,maret04,parise05,brinch09,coutens12}).
The sublimation rate of molecular ices depends exponentially on the local temperature, so a step-function is an appropriate way to model the very steep increase that occurs in gas-phase molecular abundances at specific radii \citep{rodgers03,jorgensen05}.


\floattable
\begin{deluxetable}{lcC}
\tablecaption{Parameters for \ce{H^{13}CN} emission model of \irasfull{} \label{tab:modelparams} }
\tablehead{
	\multicolumn{3}{c}{Fixed parameters} 
}
\startdata
$L_*$ $\left( L_\odot \right)$ & 22 & \\
$D$ (pc) 					   & 120 & \\
$r_{\textrm{min}}$ (AU)        & 22 & \\
$r_{\textrm{max}}$ (AU)        & 6100 & \\
$r_{\textrm{infall}}$ (AU)     & 1280 & \\
{$\rho$ power law index}      & $-1.5$ & ($r<r_{\textrm{infall}}$) \\
{                      }      & $-2$  &($r>r_{\textrm{infall}}$) \\
$M_{\textrm{env}}$ $\left(M_{\odot} \right)$ & 1.9 & \\
$M_*$ $\left(M_{\odot}\right)$ & 1.0 & \\
\hline 
\vspace{0.01cm} \\
\multicolumn{3}{c}{Floating parameters} \\
{ } & Best-fit ($\pm$) & \textrm{Allowed Range} \\
\hline
\Xin{} $\left( \ce{H^{13}CN} \right)$ & $ 8.51_{-0.74}^{+0.81} \times 10^{-10}$ & 10^{-12} - 10^{-8} \\
\Xout{} $\left( \ce{H^{13}CN} \right)$ & $ 1.82_{-0.09}^{+0.09} \times 10^{-11}$ & 10^{-14} - X_{\mathrm{in}} \\
\Tjump{} (K) & $71.2_{-2.6}^{+2.4}$ & 30 - 120 \\
\hline
\vspace{0.01cm} \\
\multicolumn{3}{c}{MCMC parameters} \\
\hline
\# walkers & 24 & \\
\# steps & 1776 & \\
Autocorrelation time $\tau$ & 34.7 & \\
\# steps / $\tau$ & 51.2 & \\
\enddata
\tablecomments{We adopt fixed physical parameters for our \iras{} model following \citet{crimier10} and \citet{coutens12}.}
\end{deluxetable}

We present the parameters used in this model in Table~\ref{tab:modelparams}.
To enable direct comparison with \citet{coutens12}, who derived the \ce{H2O} abundance using a very similar set of data and methods, we adopt the one-dimensional physical model developed by \citet{crimier10}.
This physical model incorporates (a) submillimeter single-dish emission profiles at 350, 450, and 850~$\mu$m, and (b) the spectral energy distribution (SED) from 23 to 1300~$\mu$m.
\citet{crimier10} present a model of the density and temperature structure which follows a Shu ``inside-out'' collapsing envelope \citep{shu77,adams86}, in which the density follows two power laws: $\rho \propto r^{-1.5}$ for $r<r_{\textrm{infall}}$, and $\rho \propto r^{-2}$ for $r>r_{\textrm{infall}}$.
Three-dimensional models of the physical structure of \iras{} are under development (e.g., \citealt{jacobsen17}, who model the envelope, disks and dust filament), but for the purposes of this paper, a 1D model is sufficient to derive bulk abundance properties.

In our modeling, we start with the luminosity and density distributions listed in Table~\ref{tab:modelparams}, and first solve for the temperature structure using the radiative transfer code TRANSPHERE \citep{dullemond02}\footnote{The TRANSPHERE Fortran code is hosted online at \url{http://www.ita.uni-heidelberg.de/~dullemond/software/transphere/index.shtml}}.
Following \citet{coutens12}, we use RATRAN \citep{hogerheijde00,hogerheijde00ascl,vandertak07a}, a spherical Monte Carlo 1D radiative transfer code: RATRAN solves the line radiative problem by iteratively computing the mean radiation field $J_\nu$ in each radial shell to derive the level populations of \ce{H^{13}CN}.
\changesbf{The HCN collision rate coefficients were derived by \citet{dumouchel10} with He as the collision partner. 
The rates were retrieved in a molecular data file from the LAMDA database\footnote{\url{http://home.strw.leidenuniv.nl/~moldata/}} \citep{schoier05}, where they were scaled by a factor 1.37 to represent collisions with \ce{H2}.}
Like \citet{coutens12}, we use an infalling radial velocity field $v_r = \sqrt{2GM/r}$ for a 1 $M_\odot$ central star; outside of the infall radius $r_{\textrm{infall}}$, the envelope is assumed to be static (i.e., $v_r=0$).
The output of RATRAN is a datacube (dimensions R.A., Decl., and radial velocity) of molecular line emission for each emission line under consideration.

To synthesize the observed single-dish spectra, we use the MIRIAD software package \citep{sault95} to appropriately extract information from the RATRAN-produced datacubes.
By convolving with the telescope beam profile at each frequency, and extracting the emission corresponding to the appropriate pointing offset, we produce synthetic spectra that are faithful to what would be observed with the given observational setup.
In this step, we carefully observe the different offset pointings of the ground-based TIMASSS data (pointed at \iras{}B, located 5$\arcsec$ from source A), the APEX data (pointed directly at source A), and the \textit{Herschel} data (pointed halfway between sources A and B, i.e., 2.5$\arcsec$ from source A).
Following \citet{coutens12}, we assume that our spherical model is centered on source A, the more massive of the two components.
Finally, we resample the synthetic spectra to the velocity sampling of the observed data.\footnote{RATRAN does not produce hyperfine structure for the $J=1-0$ line, as collisional rates for the separate hyperfine states of the $J=1-0$ line of \ce{H^{13}CN} are not available. 
Therefore, we simulate the hyperfine structure of the $J=1-0$ line by distributing the total flux of the $1-0$ line among the three hyperfine states according to their expected flux ratios. 
This would be inappropriate if the \ce{H^{13}CN} $J=1-0$ line were highly optically thick, but as the emprical flux ratios of the hyperfine components are very near the ideal LTE case of 1:5:3, this is a reasonable approach.}

\subsection{Fitting}

Our model fitting procedure uses a Markov chain Monte Carlo (MCMC) method.
The specific MCMC implementation we use is the ``emcee'' package \citep{foreman-mackey13,foreman-mackey13ascl}, a Python implementation of the affine-invariant MCMC sampler proposed by \citet{goodman10}.
The advantage of this MCMC approach is that it yields a posterior probability distribution of model parameters from which we can estimate uncertainties in the best-fit model parameters.
A review of Markov Chain Monte Carlo techniques and their use in astrophysics has been recently presented by \citet{hogg18}.

To compare the observed data to the model spectra, we compute the following $\chi^2$ statistic for each set of model parameters \Xin, \Xout, \Tjump:
\begin{equation}
	\chi^2 = \sum_{i=1}^{N} \sum_{j=1}^{n_\textrm{chan}} \frac { \left( T_{\textrm{data},i,j} - T_{\textrm{model},i,j}  \right)^2  } { \textrm{rms}^2_i + \left( Cal_i \times T_{\textrm{data},i,j} \right)^2 }
\end{equation}
for $N$ lines (each designated $i$), $n_\textrm{chan}$ channels per line (each designated $j$).
The observed intensity in channel $j$ of line $i$ of the data and model is designated $ T_{\textrm{data},i,j}$ and $T_{\textrm{model},i,j}$, respectively.
The per-channel rms for each line $i$ is designated $\textrm{rms}_i$.
The calibration uncertainty, denoted $Cal_i$, is fixed at 15\% for each line.

For the MCMC run, the corresponding log-likelihood function $\ln \mathscr{L}$ is written as:
\begin{equation}
	\ln \mathscr{L} = K - \frac12 \chi^2
\end{equation}
which is maximized in the MCMC fitting procedure.\footnote{As these models are highly nonlinear, we refrain from computing or reporting ``reduced chi-squared'' values for these model fits, following \citet{andrae10}.}
We constrain the model parameters (via ``flat'' priors) to the following ranges:
$X_{\textrm{in}}: 10^{-12} - 10^{-8}$, $X_{\textrm{out}}: 10^{-14} - X_{\textrm{in}}$, and $T_{\textrm{jump}}: 30 - 120$ K.

We ran the MCMC sampler with an ensemble of 24 walkers for 1776 steps.\footnote{Using a 2015 model MacBook Pro, this model took roughly 1400 s per MCMC ensemble step; the 1776-step MCMC sampler ran for 690 hours.}
\citet{foreman-mackey13} discuss ways to assess the robustness of an MCMC run, and recommend:
\begin{enumerate}
 \item that the ``acceptance fraction'' fall between $0.2-0.5$, 
 \item that the autocorrelation time $\tau$ be much less than the number of ensemble steps.
\end{enumerate}
We find the mean acceptance fraction for each of the walkers to be 0.633, which is slightly higher than the ideal range, but not enough to raise concern.
The maximum autocorrelation time $\tau$ among the three parameters was 34.7 steps; thus, we ran the MCMC chains for a factor $51 \times$ longer than $\tau$, indicating that this MCMC run has successfully converged.
Finally, we have discarded the first 69 steps (i.e., $2 \times \tau$) of the MCMC chains as ``burn-in,'' to ensure that the initial walker positions do not have an effect on the results presented below.

\subsection{Results of model fitting}
\begin{figure*}
	\plotone{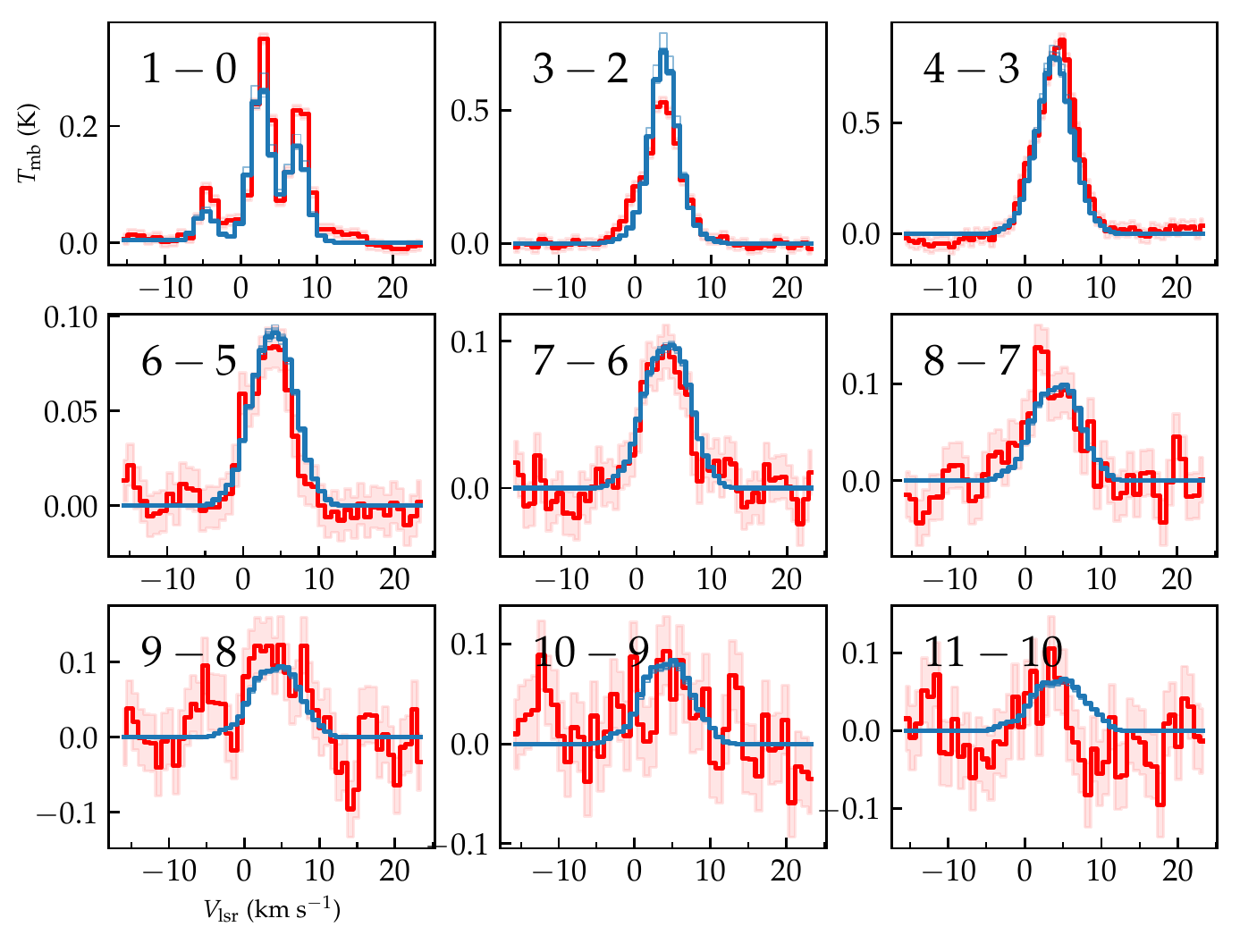}
	\caption{Comparison of observed \ce{H^{13}CN} spectra (red, with pink error bars) with ten representative model spectra (light blue lines) drawn from the posterior parameter distribution. The ``best-fit'' model (i.e., with the median values drawn from each parameter distribution) is shown with the thickest blue line. \changesbf{For most lines, the ten model spectra overlap too closely to distinguish.}\label{fig:model}}
\end{figure*}

\begin{figure*}
	\plotone{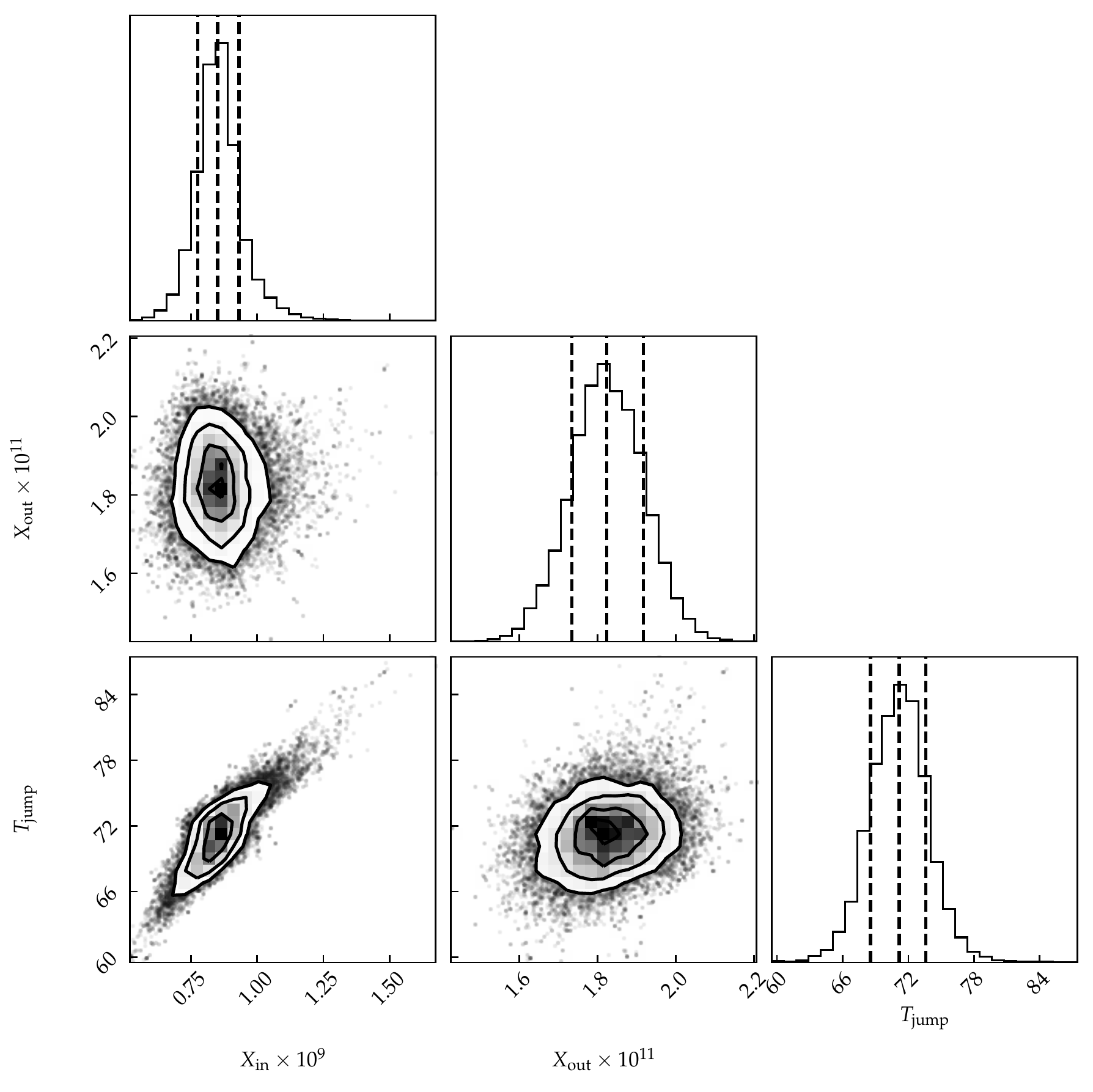}
	\caption{Corner plot of the posterior probability distributions projected into one and two dimensions along different parameter axes. Created with ``corner.py'' \citep{Foreman-Mackey:2016}. \label{fig:corner}}
\end{figure*}

We present the following ``best-fit'' (i.e. median) parameters for the posterior probability distribution of the \ce{H^{13}CN} emission model:
\begin{eqnarray*}
	X_{\mathrm{in}} (\ce{H^{13}CN}) &=& \left( 8.51_{-0.74}^{+0.81} \right) \times 10^{-10} \\
	X_{\mathrm{out}} (\ce{H^{13}CN})&=& \left( 1.82_{-0.09}^{+0.09} \right) \times 10^{-11} \\
	T_{\mathrm{jump}} &=& 71.2_{-2.6}^{+2.4}\textrm{\changesbf{ K}}
\end{eqnarray*}
and our best-fit model is shown in Figure~\ref{fig:model}, overplotted on the data from TIMASSS, APEX, and \textit{Herschel}.
Our quoted uncertainties are drawn from the 16th, 50th, and 84th percentile thresholds (i.e., the median value $\pm 34\%$) of the sample distributions on each parameter.
(This is roughly analogous to quoting $\pm 1 \sigma$ errors from a Gaussian distribution.)
We show the sample distributions for each model parameter (projected into a ``corner plot'') in Figure~\ref{fig:corner}.
From the posterior distributions, it is clear that \Xin{} and \Tjump{} are somewhat correlated, indicating the importance of varying these two parameters simultaneously, rather than assuming a \Tjump{} from the literature.
The apparently tight constraints on these parameters likely are due to the wide range of $E_{\textrm{up}}$, from $4-273$~K (cf. Table~\ref{tab:model_lines}), sampled in the nine observed lines.

By adopting a \ce{^{12}C}/\ce{^{13}C} ratio of $69 \pm 6$ \citep{wilson99}, and assuming that carbon isotopic fractionation is negligible in HCN, we therefore infer the following HCN abundances:
\begin{eqnarray*}
	X_{\mathrm{in}} (\ce{HCN}) &=& \left( 5.87_{-0.76}^{+0.72} \right) \times 10^{-8} \\
	X_{\mathrm{out}} (\ce{HCN})&=& \left( 1.26_{-0.13}^{+0.13} \right) \times 10^{-9} 
\end{eqnarray*}
\noindent We estimated the uncertainties on these parameters by summing in quadrature the $X$(\ce{H^{13}CN}) error bars with the uncertainty on the \ce{^{12}C}/\ce{^{13}C} ratio.

\subsection{Comments, limitations, and caveats of IRAS 16293 analysis}

The \Xout{} which we have derived for the outer envelope closely matches the measurement presented in \citet{schoier02}.
\citet{schoier02} use an abundance model, fit to the $J=4-3$ and $3-2$ lines of HCN and \ce{H^{13}CN}, to measure $X(\ce{H^{13}CN}) = 1.8 \times 10^{-11}$. 
They note that, as they sample only transitions which probe energy states below 90~K, they cannot constrain the existence of a hot core abundance jump for these species, so they limit their analysis to the colder outer envelope.
They derive a $X(\ce{HCN}) = 1.1 \times 10^{-9}$ (assuming $\ce{^{12}C}/\ce{^{13}C} = 60$). 
If scaled to our chosen $\ce{^{12}C}/\ce{^{13}C} = 69$ for equivalent comparison, this gives $X(\ce{HCN}) = 1.26 \times 10^{-9}$.
Their independent measurements of the outer \ce{H^{13}CN} and HCN abundances are essentially identical to our measured \Xout{}$(\ce{H^{13}CN})= \left( 1.82_{-0.09}^{+0.09} \right) \times 10^{-11}$, and $X_{\mathrm{out}} (\ce{HCN})= \left( 1.26_{-0.13}^{+0.13} \right) \times 10^{-9}$.

Our analysis, which assumes spherical symmetry of \iras{}, has known limitations, but the assumption of spherical symmetry is justifiable for our scientific purposes.
\citet{jorgensen02,jorgensen05} discuss how certain systematic effects in protostar models, such as the assumption of spherical symmetry and the uncertainty of assumed dust properties, may cause systematic uncertainties by factors of $2-3$ in absolute abundances.
As noted in depth in \citet{jorgensen16}, \iras{} is a multiple system with a non-trivial morphology on small scales.
Because we are only interested in radially averaged bulk properties of the \iras{} system, the spherically symmetric approach that we have taken is sufficient.
While there may be systematic issues in the absolute abundance scaling of our HCN measurement, the relative ratios of HCN/\ce{H2O} presented in Section~\ref{sec:discussion} use \ce{H2O} measurements from \citet{coutens12} which use the same physical model, allowing \changesbf{some} systematic issues to cancel out in the ratio.
\changesbf{One additional concern is that the $X(\ce{H2O})$ measurement from \citet{coutens12} may only be a lower limit on the water abundance, and that deviation from spherical symmetry on small scales may bias the water abundance measurement from its true value near $10^{-4}$ \citep{visser13}.
We consider this issue in Section~\ref{sec:bulk_nitrogen}.}

We omit the secondary companion, \iras B, in this model; 
meanwhile, the ground-based observation of the $J=1-0$ line from TIMASSS \citep{caux11} is centered on source B.
This is not a concern for our analysis, as previous data (especially SMA and ALMA resolved images of HCN from, e.g., \citealt{takakuwa07,zapata13}) reveal that Source A is far brighter in molecular line and continuum emission.
For the $1-0$ line, the half-power beam width of 29\arcsec{} comprises both sources well; this, combined with the fact that we have properly considered how the flux from source A would be diluted by the pointing offset of 5\arcsec{}, effectively mitigates any errors that might arise in this analysis.


\subsection{HCN sublimation temperature and binding energy}
\label{sec:binding}

Our observational coverage of nine \ce{H^{13}CN} transitions towards \iras{}, spanning $4-273$~K in upper state energy, removes our need to assume a given \Tjump{} and instead allows us to constrain it using the data. 
In the MCMC posterior distributions shown in Figure~\ref{fig:corner}, two results from the \Tjump{} posterior probability distribution are apparent:
(a) the strong positive correlation between \Tjump{} and \Xin{}, and
(b) the relatively tight constraints placed on \Tjump{} given our model assumptions and the data.
We discuss the consequences of these two results here.

The high positive correlation between \Tjump{} and \Xin{} (Pearson correlation coefficient $R=0.82$) is not surprising,
as \Tjump{} effectively sets the radius of the ``jump'' abundance zone.
For a given column density of high-temperature HCN, a higher \Tjump{} (and therefore smaller sublimation region) necessitates a higher density (and therefore abundance) of those hot HCN molecules, and vice versa.
Nonetheless, it does speak to the importance of estimating \Tjump{} accurately in order to properly measure \Xin{}.
In contrast, neither \Xin{}$-$\Xout{} ($R=0.002$) nor \Xout{}$-$\Tjump{} ($R=0.23$) show meaningful correlation.
As \Tjump{} and \Xin{} are so correlated, a change of only 20~K in the assumed \Tjump{} corresponds to a $2 \times$ difference in the derived \Xin{}.
Not all observational studies have access to \textit{Herschel} data which constrain so many transitions and energy states of an observed molecule, but when available, using these data to simultaneously constrain \Tjump{} alongside the measured abundances offers this significant advantage.

Our measurement of \Tjump{} also gives information about the binding of HCN ice onto interstellar dust grains.
As noted in \citet{hollenbach09}, there is a direct connection between the freezing (or sublimation) temperature of a molecule and the value of its binding energy to dust grains.
This energy is sometimes referred to as the adsorption or desorption energy, and denoted $E_{\textrm{B}}$, $E_{\textrm{ads}}$, or $E_{\textrm{D}}$.
The binding energy of HCN onto dust grains is poorly constrained, as laboratory measurements involving cyanide molecules (including the temperature programmed desorption experiments used to measure binding energies) are difficult to carry out safely and accurately. 
Previous estimates of HCN's binding energy range widely:
1760 K \citep{hasegawa93},
1722 K \citep{bergin95},
2050 K (\citealt{garrod06}, as listed in the fifth release of the UMIST Database for Astrochemistry, \citealt{mcelroy13}),
$3370 - 3610$ K \citep{noble13},
4700 K \citep{szőri14},
$2460 - 6974$ K \citep{bertin17}, and
3700 K (\citealt{wakelam17}, as listed in KIDA, the Kinetic Database for Astrochemistry, \citealt{wakelam12}).
This spread of binding energies corresponds to a \Tjump{} range of $30-122$~K under typical conditions.

Our derived \Tjump{} allows us to make a prediction for the binding energy of HCN onto astronomical dust grains.
By rewriting Equation~5 from \citet{hollenbach09}, which is itself derived from the Polanyi-Wigner equation \citep{polanyi25}, we can express the binding energy in terms of the jump temperature as well as several other terms:
\begin{equation}
\label{eq:binding}
\frac{E_{\textrm{ads}}(\ce{HCN})}{k} = T_{f,{\ce{HCN}}} \times \left[57 +
   \ln \left[ 
       \left(
           \frac{N_{s,{\ce{HCN}}}}{10^{15}~\mathrm{cm}^{-2}}
           \right)
       \left( 
           \frac{\nu_{\ce{HCN}}}{10^{13}~\mathrm{s}^{-1}}
           \right)
       \left(
           \frac{1~\mathrm{cm}^{-3}}{n_{\ce{HCN}}}
           \right)
       \left(
           \frac{10^{4}~\mathrm{cm~s}^{-1}}{v_{\ce{HCN}}}
           \right)
           \right] \right] 
\end{equation}
with $T_{f,\ce{HCN}}$  standing in for \Tjump{}, $N_{s,\ce{HCN}}$ the number of adsorption sites per cm$^2$, $n_{\ce{HCN}}$ the number density of HCN in the gas phase, $v_{\ce{HCN}}$ its thermal speed.
To saturate a single-molecule monolayer of ice, a gas-phase abundance of $\sim 10^{-6}$ relative to \ce{H2} is typically needed; as the HCN abundance is roughly an order of magnitude below this, we take $N_{s,\ce{HCN}} = 10^{14}~\mathrm{\changesbf{cm}}^{-2}$, i.e., $10\%$ of available sites.
We calculate the vibrational frequency $\nu_{\ce{HCN}} = 5.27 \times 10^{13}$ s$^{-1}$.
At the radius in our model at which $T$=\Tjump, the number density $n_{\ce{HCN}} = 4.9$ cm$^{-3}$, and the thermal speed $v_{\ce{HCN}} = 2.37 \times 10^4$ cm s$^{-1}$.

We predict a binding energy $E_{\textrm{ads}}(\ce{HCN})/k = 3840 \pm 140$ K,
given the derived \Tjump$ = 71.2_{-2.6}^{+2.4}$ K and the physical conditions (density, temperature, and gas-phase HCN abundance) at the corresponding radius in the envelope model.
As $E_{\textrm{ads}}$ depends linearly on \Tjump{} but only on the logarithm of the other terms in Equation~\ref{eq:binding}, we assume that uncertainties in \Tjump{} dominate in contributing to uncertainties in $E_{\textrm{ads}}$.

We note that this modeling experiment was not designed to place firm constraints on the HCN binding energy, and we have not carefully investigated the sensitivity of this prediction on the various model parameters (especially including those which we consider ``fixed'', such as the stellar luminosity and the physical envelope structure).
Careful use of this result should be observed.


\section{Discussion: On the origins of cometary nitrogen}
\label{sec:discussion}

Previous work (e.g. \citealt{Bergin15,furi15,alexander17}) has suggested that refractory carbonaceous grains and molecular ices are the precursors to the nitrogen seen in the Earth's cometary and meteoritic building blocks.
In this study we have measured the nitrogen content in organic ices towards protostars at an early stage of planet formation, which offers an opportunity to compare these measurements to a late stage of planet formation. 
Specifically, we can compare to the ice and refractory contents of solar system comets, which are considered a relatively pristine reservoir of the materials available to the young solar system during planetary assembly (e.g., \citealt{charnley08}).
Through this comparison, we can explore whether we can identify (or rule out) the original source(s) of nitrogen provided to these bodies.
In this section, we make use of two ratios to explore this topic:
the N/\ce{H2O} ratio (to trace the bulk nitrogen), 
and the \ce{^{14}N}/\ce{^{15}N} ratio (to discern between different origin populations).

The nature of nitrogen in cometary matter, as revealed by studies of material from comets 
Halley \citep{kissel87},
81P/Wild~2 (\citealt{sandford06}, and other references compiled in \citealt{alexander17}), 
and 67P/Churyumov-Gerasimenko \citep{fray16}, 
indicates that the nitrogen-bearing organic and refractory matter in comets has close isotopic and mineralogical similarities to that seen in interplanetary dust particles (IDPs) and the insoluble organic matter (IOM) in carbonaceous chondrites \citep{alexander07}. 
Further, these similarities suggest a genetic relationship between chondritic, IDP, and cometary organic matter \citep{furi15,alexander17}.


\floattable
\begin{deluxetable}{ccccc}
\tablecaption{Comparisons of N/\ce{H2O} ratios between protostars, comets, and ISM dust\label{tab:comet} }
\tablehead{
	\colhead{ } & \multicolumn{2}{c}{\textbf{Protostellar sources}\tablenotemark{a}} & \colhead{\textbf{Cometary sources}\tablenotemark{b}} & \colhead{\textbf{ISM dust}\tablenotemark{c}} 
	\\
	\colhead{ } & \colhead{\iras{}} & \colhead{\orionkl{}} & \colhead{ } & \colhead{ } 
	}
\startdata
HCN/\ce{H2O}       & $0.05-2.0\%$     & $0.10 \pm 0.06~\%$ & $0.1 - 0.6~\%$ & ... \\
organic N/\ce{H2O} & ...              & $0.13 \pm 0.07~\%$ & $0.1 - 0.9~\%$ & ... \\
\ce{NH3}/\ce{H2O}  & ...              & $0.12 - 0.92~\%$   & $0.4 - 1.8~\%$ & ... \\
total N ices       & ...              & ...                & $0.5 - 2.7~\%$ & ... \\
\hline
N/\ce{H2O} (dust)& ...                & ...                & $5 - 24~\%$    & $\lesssim 19 \pm 12~\%$ \\
\enddata
\tablecomments{All values given as percentages (\%) relative to \ce{H2O}. 
\tablenotetext{a}{Based on this work, \citet{coutens12}, \citet{neill13a}, and \citet{Crockett14b}.}
\tablenotetext{b}{Based on \citet{mumma11} and \citet{wyckoff91}.}
\tablenotetext{c}{Based on \citet{jensen07}, \citet{whittet10,whittet13}, and \citet{nieva12}.}
}
\end{deluxetable}

\subsection{Bulk nitrogen}
\label{sec:bulk_nitrogen}

Central to this investigation is the question of which primordial sources of nitrogen has a high enough abundance to account for the nitrogen content of solar system planetesimals. 
Here we consider either molecular ices or refractory dust as these primordial nitrogen reservoirs.
To compare the bulk nitrogen content of various bodies (protostars, comets, and interstellar dust), we use water (\ce{H2O}) as a standard.
As the most abundant interstellar and cometary volatile species, \ce{H2O} is measurable in all of these systems.
Comparing against an \ce{H2O} standard allows us to explore the connections among each of these stages.
In this work we have identified N/\ce{H2O} ratios in protostars, in comets, and in interstellar dust.
These are compiled in Table~\ref{tab:comet}.
While molecular ratios are commonly reported relative to \ce{H2O} in cometary ices, this work represents the first use of N/\ce{H2O} ratios to trace the bulk nitrogen content across protostars, comets, and interstellar dust.

Water abundances have been measured in these protostellar environments by \citet{neill13a} for \orionkl{}, and \citet{coutens12} for \iras{}; 
in each case, measured using nearly identical data and techniques as those used to derive HCN abundances in this work.
We focus our attention on the warm and hot ``inner'' environments of each protostellar system.
As summarized in Table~\ref{tab:comet}, the HCN/\ce{H2O} abundance in \orionkl{} is $0.10 \pm 0.06~\%$, while all organic N carriers together yield an N/\ce{H2O} ratio of  $0.13 \pm 0.07~\%$.
For \ce{NH3}, the wide abundance range allowed by \citet{gong15} gives a range of \ce{NH3}/\ce{H2O} values of $0.12-0.92\%$.

In \iras{}, we derive an HCN/\ce{H2O} ratio in the inner envelope of $1.2 \pm 0.8 \%$ 
\changesbf{
when considering the \ce{H2O} abundance from \citet{coutens12} at face value.
However, water abundances measured in the inner envelopes of protostars may suffer from systematic issues related to the assumption of spherical symmetry, as noted in depth by \citet{visser13}, who studied \ce{H2O} in the Class~0 protostar NGC1333~IRAS2A.
\citeauthor{visser13} find that a spherically symmetric model for the inner 100~AU fails to reproduce the observed emission from many lines, and that a disk or proto-disk is needed to account for the discrepancy;
one implication of this geometric modification is a substantially higher water abundance, likely near $10^{-4}$ relative to \ce{H2}, as predicted by some theoretical studies such as \citet{rodgers03}.
The comparison of ALMA data between NGC1333-IRAS2A and \iras{} indicates that \iras{} itself has a 30-times-higher \ce{H2O} column density on scales revealed by ALMA \citep{persson13,visser13}, indicating that its abundances are also likely affected by this issue.
These geometric effects are most severe at the smallest scales, where the sublimation zone of \ce{H2O} lies, so \ce{H2O} is more affected by this issue than HCN.
We therefore consider the warm inner \ce{H2O} abundance of \iras{} to be somewhere in the range of $5 \times 10^{-6} - 10^{-4}$; when uncertainties in the HCN abundance are folded in, this yields an HCN/\ce{H2O} ratio $0.05-2.0\%$, with some preference for the lower end of this range.
}

The cometary abundances of many molecular species are compiled in \citet{mumma11} relative to \ce{H2O}.
The most abundant and second-most-abundant nitrogen-bearing ices in comets are \ce{NH3} and HCN, respectively \citep{mumma11}.
Cometary \ce{NH3} ice has an abundance of $\sim 0.4 - 1.8 \%$ relative to \ce{H2O} ice.
HCN, the next largest contributor, has an abundance of $0.1 - 0.6 \%$ relative to \ce{H2O}.
Other, less-abundant species sum to at most $0.3 \%$, giving a total N abundance in ices of $0.5 - 2.7 \%$.
We find that the protostellar abundances for molecular N-bearing ices are of roughly the same order of magnitude as the cometary ice abundances.
This suggests that the nitrogen-bearing molecular ices already present on dust grain surfaces in the protostellar stage may be the direct progenitors to the nitrogenic molecular ices found in comets, as suggested previously by numerous other studies \citep[cf.][and references therein]{mumma11}.

Only a small portion of cometary nitrogen is contained in molecular ices, however.
\citet{wyckoff91} measured the ratio of nitrogen in the dust to gas in Comet 1P/Halley to be 90:10, i.e., only 10\% of Halley's nitrogen is in the gas, with the vast majority in dust.
Measurements by the Rosetta probe indicate that Comet~67P also has much more N in dust:
results from COSAC and Philae \citep{Goesmannaab0689,wright15}, COSIMA \citep{fray17}, and ROSINA \citep{leroy15} all confirm that the dust of 67P is more N-rich than the gas.
Based on these studies of comets Halley and 67P, we therefore estimate the total nitrogen content in the typical comet as compiled by \citet{mumma11}, by multiplying the range of N-ices/\ce{H2O} abundances by 10$\times$.
We estimate that the N/\ce{H2O} in dust in comets is $5-24\%$, and the bulk N/\ce{H2O} ratio in comets is $5-27\%$.
Therefore, the nitrogen in molecular ices measured in protostars is an order of magnitude or more too low to account for the total nitrogen content of comets.

In principle, the most volatile forms of nitrogen (especially \ce{N2}) should contain the majority of interstellar nitrogen at the beginning of the planet formation process (see, e.g., \citealt{schwarz14}).
However, we can exclude these more volatile forms of nitrogen as contributors to cometary nitrogen, based on chemical and physical principles. 
It would be very surprising if these species were able to undergo solid-state chemistry and contribute to chemical complexity, given their low sublimation temperatures and binding energies (as shown in Table~\ref{tab:bind}) and the slow rates at which gas-phase chemistry proceeds.
The molecular species NO and NS also likely have binding energies too low to permit freeze-out and its subsequent chemical enrichment.


\floattable
\begin{deluxetable}{lLL}
\tablecaption{Binding energies and sublimation temperatures for simple N-bearing species \label{tab:bind}}
\tablehead{
	\colhead{Species} & 
	\colhead{$E_{\textrm{B}}/k$ (K)} &
	\colhead{\Tjump{} (K)}
}
\startdata
N       & 720 \pm 216  & 13 \pm 4 \\
\ce{N2} & 1100 \pm 330  & 19 \pm 6 \\
NO      & 1600 \pm 480  & 28 \pm 8 \\
NS      & 1900 \tablenotemark{a}   & 33  \\
HCN     & 3700 \pm 1100  & 65 \pm 19 \\
{ }     & \textrm{or }$3840\pm140$\tablenotemark{b}  & \textrm{or }$70 \pm 2$  \\
\ce{NH3}& 5500 \pm 1650 & 96 \pm 29 \\
\enddata
\tablenotetext{a}{\citet{garrod06}, as listed in the KIDA database \citep{wakelam12}}
\tablenotetext{b}{This work; see \S \ref{sec:binding}}
\tablecomments{Sublimation temperatures \Tjump{} calculated using typical ISM conditions following \citet{hollenbach09}. 
All binding energies from \citet{wakelam17}, as listed in the KIDA database \citep{wakelam12}, unless otherwise noted. In all cases, the binding surface is assumed to be \ce{H2O} ice.}
\end{deluxetable}

An additional source of nitrogen for comets may exist: refractory nitrogen within interstellar dust.
Nitrogen has not been detected directly in ISM dust, and it is difficult to observationally constrain the presence of nitrogen in carbon-rich dust via spectroscopic observations \citep{jones16a}. 
Nonetheless, an upper limit on nitrogen's abundance can be inferred from measurements of N in interstellar gas, specifically via the depletion of N in diffuse ISM gas relative to the cosmic abundance of nitrogen atoms.
\citet{nieva12} measure the present-day cosmic abundance of nitrogen (relative to hydrogen) in the local Universe to be N/H = $62 \pm 6$ ppm.
\citet{jensen07} measure the abundance of N~I gas along multiple sight lines of the diffuse ISM to be $51\pm4$ ppm.
Based on these numbers, the nitrogen depletion in the diffuse ISM is $11 \pm 7$~ppm.
This number may represent depletion into different types of nitrogen sinks, such as formation of \ce{N2} molecules or other species, but this depletion is measured in low-extinction (low $A_V$) environments where molecular \ce{N2} would be rapidly destroyed by interstellar ultraviolet radiation.
Therefore, the most likely interpretation is nitrogen depletion into refractory dust grains.
In any case, the $11 \pm 7$~ppm is an upper limit on the available nitrogen in interstellar refractory dust.
\citet{whittet10} and \citet{whittet13} show that the abundance of oxygen atoms within ices on interstellar dust grains is 116~ppm (relative to H), and the solar O/H abundance is 457~ppm. 
Of this, about 50$-$60\% of the oxygen atoms are in \ce{H2O}, with the rest in CO, and \ce{CO2}).
Taking 50\% of the 116~ppm as \ce{H2O} ice yields an \ce{H2O} ice-on-dust abundance of 58~ppm versus H in the ISM.
Dividing the refractory N value by this \ce{H2O} on dust, we present an upper limit of $19 \pm 12 \%$ on refractory nitrogen in the ISM relative to water.
This upper limit is comparable to the bulk nitrogen abundance in comets.
Because there are no other apparent sources of nitrogen to comets with high enough abundance, we suggest that interstellar dust is the likely origin of the majority of cometary nitrogen.

\subsection{\ce{^{15}N} enrichment}

In the previous subsection we suggested that the bulk of cometary nitrogen may be inherited from interstellar dust, as molecular ices do not have a high enough abundance relative to water to provide the majority of cometary nitrogen.
To further explore the origins of cometary nitrogen, we turn to the isotopic signature of \ce{^{15}N}.
Isotopic ratios such as the \ce{^{14}N}/\ce{^{15}N} ratio are commonly used to trace populations of material, as the isotopic ratio can be robust to physical processes and persist through time.
In the solar system, the nitrogen in planetesimals is significantly more \ce{^{15}N}-rich than the Sun and the local interstellar medium: 
the solar value \ce{^{14}N/^{15}N} ratio is 441,
while planetesimals have values ranging from 200, 50, or 140  
\citep[for bulk chondrites, chondrite hot spots, and comets, respectively]{furi15}.
Direct measurements of \Nratio{} in the dust particles of Comet~81P/Wild~2 show a range of \Nratio{} values from $\sim 180-270$ \citep{mckeegan06}.
The present-day Earth itself has a sub-solar \Nratio{} ratio of 272 \citep{anders89}.

The common interpretation of the \ce{^{15}N} enrichment in comets and other bodies is that it originates in the low-temperature molecular chemistry that occurs in either interstellar clouds or the early phases of star and planet formation \citep{charnley08,wirstrom12,hily-blant13,furi15,bockelee-morvan15,alexander17}.
\ce{^{15}N} fractionation is 
consistently observed in N-bearing molecular ices (amines \& nitriles) in dense star-forming gas \citep[][as well as this study]{hily-blant13,wampfler14}.
\citet{hily-blant17} present evidence for multiple nitrogen reservoirs within forming solar systems, with distinct isotopic signatures.
The similar N/\ce{H2O} abundances for N-bearing molecules around protostars derived in this work, and in cometary ices, further lends support to the interpretation that ices are the \ce{^{15}N} donor to comets.
However, recent work by \citet{roueff15} and \citet{wirstrom18} indicates that, when updated reaction rates and more sophisticated quantum-chemical computations are included, the current chemical networks cannot reproduce the observed \ce{^{15}N} enrichments in several N-bearing molecules, leaving room for exploration of the origins of \ce{^{15}N} in the dense ISM.
Regardless, observations clearly indicate that ISM chemistry can produce \ce{^{15}N} enrichments in distinct chemical families (e.g. nitriles).

Thus, while substantial evidence supports a molecular ice origin for \ce{^{15}N} in comets, we ask whether a dust origin be ruled out entirely.
This question is especially important to resolve given the large contribution of N from dust,
as a small \ce{^{15}N} contribution from dust would matter more than a large \ce{^{15}N} contribution from ices.
Unfortunately, the \ce{^{15}N} content of ISM dust is both unknown and totally unconstrained by observation;
this requires us to rely on indirect evidence to investigate where cometary \ce{^{15}N} originates.
Most of the nitrogen in interstellar dust would be expected to be contained in carbonaceous (rather than silicate) grains.
The formation of carbonaceous dust, discussed by \citet{chiar13}, occurs in $\sim 1000$~K environments that do not encourage chemical fractionation of nitrogen.
The incorporation of nitrogen into hydrocarbon dust, discussed by \citet{jones16a}, also would not be expected to enrich \ce{^{15}N} over \ce{^{14}N}.
The cores of interstellar dust grains consist of presolar stardust that form directly from stellar ejecta \citep{clayton04,chiar13}.
While the \Nratio{} ratio of interstellar dust is unknown, individual presolar stardust grains can survive planetesimal formation intact and are amenable to isotopic study.
These stardust grains, by virtue of their localized formation, inherit isotopic ratios directly from the nucleosynthetic and stellar evolutionary processes in their parent star.
Presolar stardust grains, particularly SiC-X, graphite, and \ce{Si3N4}, were studied by \citet{clayton04}.
Graphite grains typically have the same nitrogen isotopic ratio as the Sun, but SiC-X and \ce{Si3N4} grains are \ce{^{15}N}-rich, with \Nratio{} values as low as 20 in some cases.
\citet{clayton04} discuss how these high \ce{^{15}N} values may be linked to the dust formation and nucleosynthesis processes in supernovae.
In a related process, nova eruptions may be partly responsible for the gradual rise of \ce{^{15}N} enrichment in a galaxy over time \citep{romano17}.

Additional evidence comes from analysis of organic matter within carbonaceous chondrites. 
Notably, the \ce{^{15}N} hotspots in chondrites are not typically correlated with mineral rims or individual silicate grains \citep{alexander17}, although an instance of \ce{^{15}N} enrichment near a supernova silicate grain in an IDP was noted by \citet{messenger05}.
The variable \ce{^{15}N} enrichment (concentrated in \ce{^{15}N}-rich ``hotspots'') might be consistent with an origin in diverse, heterogeneous stardust grains which each carry a different degree of \ce{^{15}N} enrichment.
This may also be consistent with how D and \ce{^{15}N} hotspots do not exactly correlate: \ce{^{15}N} hotspots are often associated with \ce{^{13}C} isotopic anomalies, while D hotspots do not co-vary with \ce{^{13}C} \citep{alexander17}.
These isotopic co-variations may also just be the result of chemistry: \citet{wirstrom12} note that for HCN and HNC, the reactions that lead to \ce{^{15}N} enrichment do not correlate with the most extreme D enrichment.

If there exist thermal, aqueous, or chemical processes that can transport \ce{^{15}N} out of stardust grains and into organic matter during the formation of comets and other bodies,
then we may not yet be able to rule out a dust origin for \ce{^{15}N} in these planetesimal bodies.
Otherwise, most of the available evidence is in favor of a molecular ice origin, although recent work in chemical models indicates that further theoretical and laboratory work is needed to understand the precise reactions that lead to \ce{^{15}N} enrichment.
In this framework, molecular ices are the donor of \ce{^{15}N} to protoplanetary solid bodies --- and  ultimately to the Earth.
Thus, following these molecular ices allows us to astronomically trace a meaningful component of the nitrogen that later becomes a part of terrestrial worlds.

\section{Conclusions}
\label{sec:conclusions}

From our analysis of the nitrogen-bearing organic molecular inventories in a high-mass hot core, the HCN lines towards a low-mass protostar, and through scaling molecular abundances relative to \ce{H2O}, we present the following conclusions:

\begin{enumerate}

	\item HCN is by far the most abundant nitrogen-bearing organic in the \oklhc{}, carrying $74_{-9}^{+5}\%$ of nitrogen-in-organics.
	\item The HCN abundance in the envelope of \irasfull{} exhibits a jump profile, with \Xin{}$ = 5.9\pm0.7 \times 10^{-8}$ and an outer HCN abundance \Xout{}$ = 1.26 \pm 0.13 \times 10^{-9}$.
	\item We derive an HCN sublimation temperature \Tjump$ = 71 \pm 3$~K, from which we make an astronomically-motivated prediction that the HCN binding energy $E_{\textrm{B}}/k = 3840 \pm 140$~K.

	\item The N/\ce{H2O} ratio in molecular ices seen in the inner protostellar envelopes is similar to N/\ce{H2O} in cometary ices. 
	However, it is not high enough to account for the total N/\ce{H2O} seen in comets. 
	While the refractory nitrogen content in interstellar dust has not been measured, its upper limit is permissive enough to account for the bulk of cometary N.
	Therefore, we suggest that the nitrogen contained in interstellar dust is the likely precursor to most of the N in comets.

	\item The high \ce{^{15}N} enrichment seen in cometary and meteoritic bodies has an unknown origin.
	Most evidence indicates that it is donated by molecular ices that underwent chemical fractionation of nitrogen isotopes, but the reactions responsible for this process are unclear.
	A dust origin of \ce{^{15}N} enrichment appears unlikely but cannot be ruled out.

\end{enumerate}

\acknowledgments

We thank K. Villalon for assistance with the geochemistry literature.
We thank L. I. Cleeves, K. Schwarz, and S. Manigand for guidance with the molecular adsorption energy literature.
We thank L. E. Kristensen for discussions about data handling and model fitting.
We thank S. Massalkhi for assistance with the literature on giant stars.
We thank P. Hily-Blant for discussions on the origins of nitrogen.

This material is based upon work supported by the National Science Foundation
Graduate Research Fellowship Program under Grant No. 1256260 and an
international travel allowance through the Graduate Research Opportunities Worldwide
(GROW). 
This work was additionally supported by funding from NSF grants AST-1514670 as well as NASA NNX16AB48G.
Any opinions, findings, and conclusions or recommendations expressed in this
material are those of the author(s) and do not necessarily reflect the views of the National
Science Foundation. 
JKJ is supported by the European Research Council (ERC) under the European Union's Horizon 2020 research and innovation programme (grant agreement No~646908) through ERC Consolidator Grant ``S4F''. Research at Centre for Star and Planet Formation is funded by the Danish National Research Foundation.

\software{
This research made use of the following: 
Astropy, a community-developed core Python package for Astronomy \citep{Astropy13,astropy18};
GILDAS, a collection of software to reduce and analyze (sub)millimeter data (\citealt{pety05,gildas13,pety18}, \url{http://www.iram.fr/IRAMFR/GILDAS});
Matplotlib \citep{Hunter:2007}, a Python package for generating scientific plots; 
``emcee'' \citep{foreman-mackey13ascl}, a Python MCMC affine-invariant sampler;
``corner.py,'' a Python tool to visualize corner plots of MCMC posterior distributions \citep{Foreman-Mackey:2016};  and 
WebPlotDigitizer \citep{ankit_rohatgi_2017_802310}, to better access published information in the literature.
}

\appendix

\section{\textit{Herschel} Line Fitting Results for \iras{}}
\label{sec:appendix_A}

\begin{figure}
\plotone{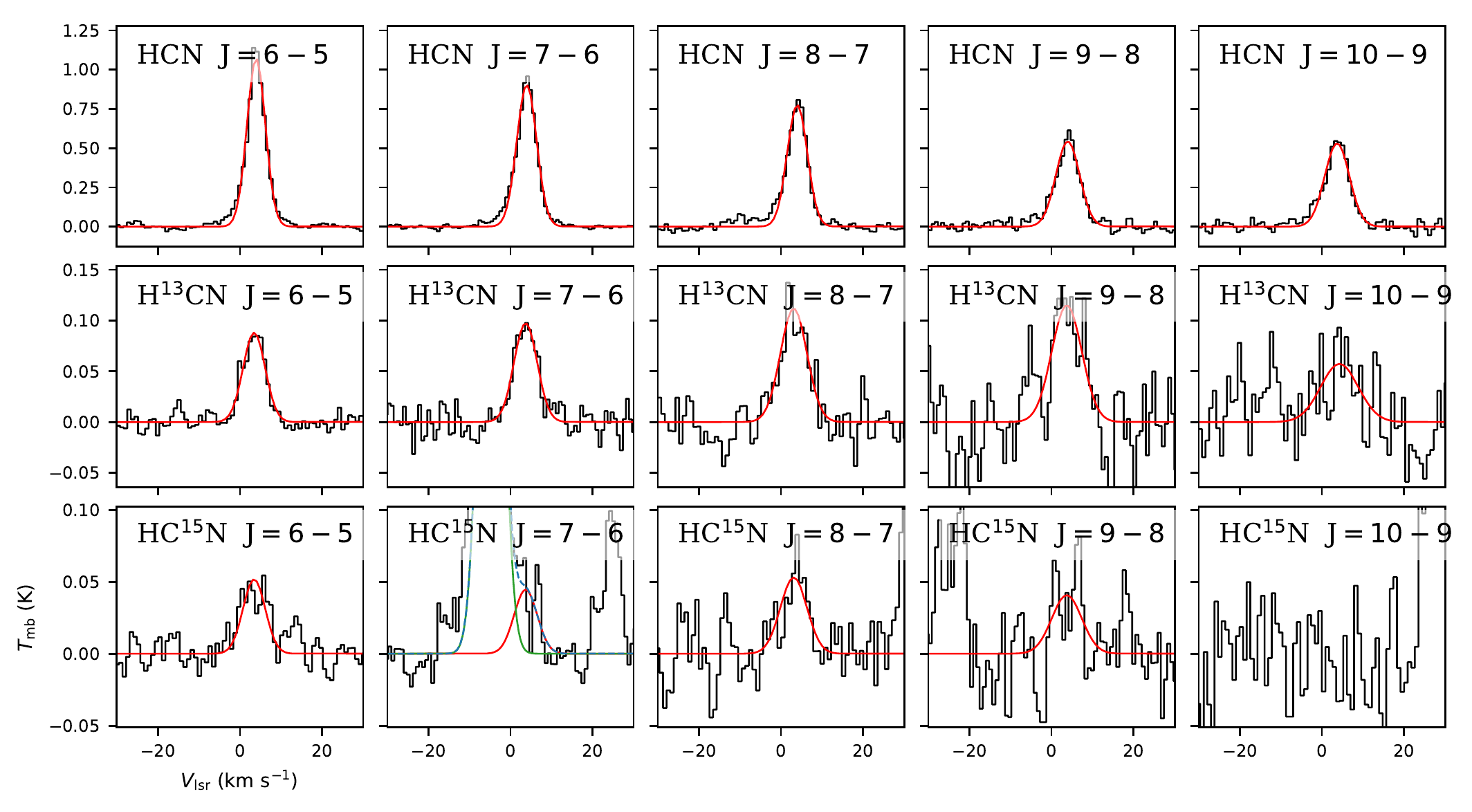}
\caption{Gaussian fits to the HCN and isotopologue lines in the HIFI data of \iras{}. Smoothed to 3$\times$ the resolution of the lowest \ce{HCN} line. 
We fit the \ce{HC^{15}N} $J=7-6$ line simultaneously with the blended SO $J=14-13$ line at 602.292~GHz.
We consider \ce{HC^{15}N} $J=10-9$ to be undetected.
\label{fig:hcn_hifi_lines}}
\end{figure}

In this Appendix, we show the detailed results of fitting the HCN, \ce{H^{13}CN}, and \ce{HC^{15}N} lines of \iras{} from the \textit{Herschel} data, as well as the derived properties (such as the optical depth $\tau$) of each state that are measured from comparisons of the isotopologues. 
The lines and their Gaussian fits are shown in Figure~\ref{fig:hcn_hifi_lines}.
The fit parameters for the \textit{Herschel} lines, as well as for the ground-based lines from \citet{caux11} and \citet{wampfler14}, are presented in \ref{tab:hcn_lines}, and the properties derived from these fits are shown in \ref{tab:hcn_derived_props}.

\begin{deluxetable}{l|ccccccc}
\tabletypesize{\scriptsize}
\tablecaption{Observed lines of HCN and isotopologues towards \iras{} \label{tab:hcn_lines}}
\tablehead{
\colhead{Species \& Transition} & \colhead{Frequency} & \colhead{Telescope} & \colhead{$\theta_{\rm{mb}}$} & \colhead{$V_0$} & \colhead{FWHM} & \colhead{peak flux} & \colhead{$\int T_{\rm{mb}} dv$} \\ 
\colhead{} & \colhead{(GHz)} & \colhead{\& Band} & \colhead{$(\arcsec)$} & \colhead{(km s$^{-1}$)} & \colhead{(km s$^{-1}$)} & \colhead{(K)} & \colhead{(K km s$^{-1}$)}
}

\startdata
HCN  J$=1-0$ & 88.6316 & IRAM-30m & 28.4 & 4.46 $\pm$ 0.33 & 9.50 $\pm$ 0.79 & 1.765 & 21.71 $\pm$ 2.39 \\
HCN  J$=3-2$ & 265.8864 & APEX-1  & 24.3 & 3.72 $\pm$ 0.01 & 6.81 $\pm$ 0.02 & 9.66 & 27.52 $\pm$ 2.3  \\
HCN  J$=6-5$  & 531.7164 & HIFI 1a & 40.5 & 3.94 $\pm$ 0.02 & 5.30 $\pm$ 0.07 & 1.072 & 6.04 $\pm$ 0.06 \\
HCN  J$=7-6$  & 620.3041 & HIFI 1b & 34.7 & 3.96 $\pm$ 0.02 & 5.70 $\pm$ 0.05 & 0.903 & 5.48 $\pm$ 0.04 \\
HCN  J$=8-7$  & 708.8772 & HIFI 2a & 30.4 & 3.98 $\pm$ 0.06 & 5.74 $\pm$ 0.15 & 0.776 & 4.74 $\pm$ 0.10 \\
HCN  J$=9-8$  & 797.4337 & HIFI 2b & 27.0 & 3.97 $\pm$ 0.09 & 6.80 $\pm$ 0.22 & 0.543 & 3.93 $\pm$ 0.11 \\
HCN  J$=10-9$ & 885.9714 & HIFI 3b & 24.3 & 3.69 $\pm$ 0.09 & 7.06 $\pm$ 0.24 & 0.531 & 3.99 $\pm$ 0.11 \\
H$^{13}$CN  J$=1-0$ & 86.3402 & IRAM-30m & 29.1 & 3.88 $\pm$ 0.09 & 2.91 $\pm$ 0.25 & 0.284 & 1.04 $\pm$ 0.12 \\
H$^{13}$CN  J$=3-2$ & 259.0118 & APEX-1  & 24.3 & 3.75 $\pm$ 0.03 & 5.70 $\pm$ 0.08 & 0.68 & 3.24 $\pm$ 0.3  \\
H$^{13}$CN  J$=4-3$ & 345.3397 & APEX-2  & 18.2 & 4.28 $\pm$ 0.04 & 6.11 $\pm$ 0.09 & 1.08 & 5.56 $\pm$ 0.8  \\
H$^{13}$CN  J$=6-5$  & 517.9698 & HIFI 1a & 41.6 & 3.44 $\pm$ 0.14 & 6.35 $\pm$ 0.30 & 0.088 & 0.60 $\pm$ 0.03 \\
H$^{13}$CN  J$=7-6$  & 604.2679 & HIFI 1b & 35.7 & 3.69 $\pm$ 0.20 & 6.62 $\pm$ 0.49 & 0.097 & 0.69 $\pm$ 0.04 \\
H$^{13}$CN  J$=8-7$  & 690.5521 & HIFI 2a & 31.2 & 3.10 $\pm$ 0.29 & 7.63 $\pm$ 0.69 & 0.112 & 0.91 $\pm$ 0.07 \\
H$^{13}$CN  J$=9-8$  & 776.8203 & HIFI 2b & 27.7 & 3.73 $\pm$ 0.61 & 8.60 $\pm$ 1.48 & 0.115 & 1.05 $\pm$ 0.15 \\
H$^{13}$CN  J$=10-9$ & 863.0706 & HIFI 3b & 25.0 & 4.27 $\pm$ 1.41 & 10.80 $\pm$ 3.91 & 0.057 & 0.66 $\pm$ 0.18 \\
HC$^{15}$N  J$=1-0$ & 86.0550 & IRAM-30m & 29.2 & 4.03 $\pm$ 0.18 & 3.51 $\pm$ 0.43 & 0.114 & 0.51 $\pm$ 0.06 \\
HC$^{15}$N  J$=3-2$ & 258.1570 & APEX-1  & 24.3 & 3.71 $\pm$ 0.09 & 6.2  $\pm$ 0.2 & 0.27 & 1.37 $\pm$ 0.1  \\
HC$^{15}$N  J$=4-3$ & 344.2001 & APEX-2  & 18.2 & 3.11 $\pm$ 0.09 & 6.4  $\pm$ 0.2 & 0.44 & 2.02 $\pm$ 0.3  \\
HC$^{15}$N  J$=6-5$  & 516.2615 & HIFI 1a & 41.8 & 3.44 $\pm$ 0.14 & 6.35 $\pm$ 0.30 & 0.052 & 0.35 $\pm$ 0.04 \\
HC$^{15}$N  J$=7-6$  & 602.2754 & HIFI 1b & 35.8 & 3.69 $\pm$ 0.20 & 6.62 $\pm$ 0.49 & 0.044 & 0.31 $\pm$ 0.08 \\
HC$^{15}$N  J$=8-7$  & 688.2758 & HIFI 2a & 31.3 & 3.10 $\pm$ 0.29 & 7.63 $\pm$ 0.69 & 0.053 & 0.43 $\pm$ 0.25 \\
HC$^{15}$N  J$=9-8$  & 774.2605 & HIFI 2b & 27.8 & 3.73 $\pm$ 0.61 & 8.60 $\pm$ 1.48 & 0.041 & 0.37 $\pm$ 0.11 \\
\enddata
\end{deluxetable}


\begin{deluxetable}{l|cc|ccc}
\tablecaption{Derived properties of each energy state of HCN towards \iras{}\label{tab:hcn_derived_props}}
\tablehead{
	\colhead{$J_u$} & 	\colhead{$\tau_{\ce{H^{13}CN}}$} & \colhead{$\tau_{\ce{HCN}}$} &  
	\colhead{$\tau_{\ce{HC^{15}N}}$} & 	\colhead{\ce{H^{13}CN / HC^{15}N}} & \colhead{\ce{^{14}N / ^{15}N}} }
\startdata
6 & 0.09 & 5.92 & 0.05 & 1.7 & 119 \\
7 & 0.11 & 7.87 & 0.05 & 2.3 & 155 \\
8 & 0.16 & 10.73 & 0.07 & 2.2 & 151 \\
9 & 0.24 & 16.41 & 0.08 & 3.0 & 210 \\
10 & 0.11 & 7.87 & --- & --- & ---
\enddata
\tablecomments{Derived properties for HCN and \ce{^{14}N / ^{15}N} assume a \ce{^{12}C / ^{13}C} ratio of $69\pm6$ following \citet{wilson99}.}
\end{deluxetable}









\bibliography{bibliography}

\begin{thebibliography}{}
\expandafter\ifx\csname natexlab\endcsname\relax\def\natexlab#1{#1}\fi

\bibitem[{{Adams} \& {Shu}(1986)}]{adams86}
{Adams}, F.~C., \& {Shu}, F.~H. 1986, \apj, 308, 836

\bibitem[{{Adande} {et~al.}(2013){Adande}, {Woolf}, \& {Ziurys}}]{adande13}
{Adande}, G.~R., {Woolf}, N.~J., \& {Ziurys}, L.~M. 2013, Astrobiology, 13, 439

\bibitem[{{Alexander} {et~al.}(2012){Alexander}, {Bowden}, {Fogel}, {Howard},
  {Herd}, \& {Nittler}}]{alexander12}
{Alexander}, C.~M.~O.~., {Bowden}, R., {Fogel}, M.~L., {et~al.} 2012, Science,
  337, 721

\bibitem[{{Alexander} {et~al.}(2017){Alexander}, {Cody}, {De Gregorio},
  {Nittler}, \& {Stroud}}]{alexander17}
{Alexander}, C.~M.~O.~., {Cody}, G.~D., {De Gregorio}, B.~T., {Nittler}, L.~R.,
  \& {Stroud}, R.~M. 2017, Chemie der Erde / Geochemistry, 77, 227

\bibitem[{{Alexander} {et~al.}(2007){Alexander}, {Fogel}, {Yabuta}, \&
  {Cody}}]{alexander07}
{Alexander}, C.~M.~O.~., {Fogel}, M., {Yabuta}, H., \& {Cody}, G.~D. 2007,
  \gca, 71, 4380

\bibitem[{{Anders} \& {Grevesse}(1989)}]{anders89}
{Anders}, E., \& {Grevesse}, N. 1989, \gca, 53, 197

\bibitem[{{Andrae} {et~al.}(2010){Andrae}, {Schulze-Hartung}, \&
  {Melchior}}]{andrae10}
{Andrae}, R., {Schulze-Hartung}, T., \& {Melchior}, P. 2010, ArXiv e-prints,
  arXiv:1012.3754

\bibitem[{{Astropy Collaboration} {et~al.}(2013){Astropy Collaboration},
  {Robitaille}, {Tollerud}, {Greenfield}, {Droettboom}, {Bray}, {Aldcroft},
  {Davis}, {Ginsburg}, {Price-Whelan}, {Kerzendorf}, {Conley}, {Crighton},
  {Barbary}, {Muna}, {Ferguson}, {Grollier}, {Parikh}, {Nair}, {Unther},
  {Deil}, {Woillez}, {Conseil}, {Kramer}, {Turner}, {Singer}, {Fox}, {Weaver},
  {Zabalza}, {Edwards}, {Azalee Bostroem}, {Burke}, {Casey}, {Crawford},
  {Dencheva}, {Ely}, {Jenness}, {Labrie}, {Lim}, {Pierfederici}, {Pontzen},
  {Ptak}, {Refsdal}, {Servillat}, \& {Streicher}}]{Astropy13}
{Astropy Collaboration}, {Robitaille}, T.~P., {Tollerud}, E.~J., {et~al.} 2013,
  \aap, 558, A33

\bibitem[{{Bacmann} {et~al.}(2010){Bacmann}, {Caux}, {Hily-Blant}, {Parise},
  {Pagani}, {Bottinelli}, {Maret}, {Vastel}, {Ceccarelli}, {Cernicharo},
  {Henning}, {Castets}, {Coutens}, {Bergin}, {Blake}, {Crimier}, {Demyk},
  {Dominik}, {Gerin}, {Hennebelle}, {Kahane}, {Klotz}, {Melnick}, {Schilke},
  {Wakelam}, {Walters}, {Baudry}, {Bell}, {Benedettini}, {Boogert}, {Cabrit},
  {Caselli}, {Codella}, {Comito}, {Encrenaz}, {Falgarone}, {Fuente},
  {Goldsmith}, {Helmich}, {Herbst}, {Jacq}, {Kama}, {Langer}, {Lefloch}, {Lis},
  {Lord}, {Lorenzani}, {Neufeld}, {Nisini}, {Pacheco}, {Pearson}, {Phillips},
  {Salez}, {Saraceno}, {Schuster}, {Tielens}, {van der Tak}, {van der Wiel},
  {Viti}, {Wyrowski}, {Yorke}, {Faure}, {Benz}, {Coeur-Joly}, {Cros},
  {G{\"u}sten}, \& {Ravera}}]{bacmann10}
{Bacmann}, A., {Caux}, E., {Hily-Blant}, P., {et~al.} 2010, \aap, 521, L42

\bibitem[{Bergin {et~al.}(2015)Bergin, Blake, Ciesla, Hirschmann, \&
  Li}]{Bergin15}
Bergin, E.~A., Blake, G.~A., Ciesla, F., Hirschmann, M.~M., \& Li, J. 2015,
  Proceedings of the National Academy of Sciences, 112, 8965

\bibitem[{{Bergin} {et~al.}(2014){Bergin}, {Cleeves}, {Crockett}, \&
  {Blake}}]{bergin14}
{Bergin}, E.~A., {Cleeves}, L.~I., {Crockett}, N., \& {Blake}, G.~A. 2014,
  Faraday Discussions, 168, arXiv:1405.7394

\bibitem[{{Bergin} {et~al.}(1995){Bergin}, {Langer}, \& {Goldsmith}}]{bergin95}
{Bergin}, E.~A., {Langer}, W.~D., \& {Goldsmith}, P.~F. 1995, \apj, 441, 222

\bibitem[{{Bergin} {et~al.}(2010){Bergin}, {Phillips}, {Comito}, {Crockett},
  {Lis}, {Schilke}, {Wang}, {Bell}, {Blake}, {Bumble}, {Caux}, {Cabrit},
  {Ceccarelli}, {Cernicharo}, {Daniel}, {de Graauw}, {Dubernet},
  {Emprechtinger}, {Encrenaz}, {Falgarone}, {Gerin}, {Giesen}, {Goicoechea},
  {Goldsmith}, {Gupta}, {Hartogh}, {Helmich}, {Herbst}, {Joblin}, {Johnstone},
  {Kawamura}, {Langer}, {Latter}, {Lord}, {Maret}, {Martin}, {Melnick},
  {Menten}, {Morris}, {M{\"u}ller}, {Murphy}, {Neufeld}, {Ossenkopf}, {Pagani},
  {Pearson}, {P{\'e}rault}, {Plume}, {Roelfsema}, {Qin}, {Salez}, {Schlemmer},
  {Stutzki}, {Tielens}, {Trappe}, {van der Tak}, {Vastel}, {Yorke}, {Yu}, \&
  {Zmuidzinas}}]{bergin10}
{Bergin}, E.~A., {Phillips}, T.~G., {Comito}, C., {et~al.} 2010, \aap, 521, L20

\bibitem[{{Bertin} {et~al.}(2017){Bertin}, {Doronin}, {Fillion}, {Michaut},
  {Philippe}, {Lattelais}, {Markovits}, {Pauzat}, {Ellinger}, \&
  {Guillemin}}]{bertin17}
{Bertin}, M., {Doronin}, M., {Fillion}, J.-H., {et~al.} 2017, \aap, 598, A18

\bibitem[{{Bockel{\'e}e-Morvan} {et~al.}(2015){Bockel{\'e}e-Morvan},
  {Calmonte}, {Charnley}, {Duprat}, {Engrand}, {Gicquel}, {H{\"a}ssig},
  {Jehin}, {Kawakita}, {Marty}, {Milam}, {Morse}, {Rousselot}, {Sheridan}, \&
  {Wirstr{\"o}m}}]{bockelee-morvan15}
{Bockel{\'e}e-Morvan}, D., {Calmonte}, U., {Charnley}, S., {et~al.} 2015, \ssr,
  197, 47

\bibitem[{{Bonal} {et~al.}(2010){Bonal}, {Huss}, {Krot}, {Nagashima}, {Ishii},
  \& {Bradley}}]{bonal10}
{Bonal}, L., {Huss}, G.~R., {Krot}, A.~N., {et~al.} 2010, \gca, 74, 6590

\bibitem[{{Briani} {et~al.}(2009){Briani}, {Gounelle}, {Marrocchi},
  {Mostefaoui}, {Leroux}, {Quirico}, \& {Meibom}}]{briani09}
{Briani}, G., {Gounelle}, M., {Marrocchi}, Y., {et~al.} 2009, Proceedings of
  the National Academy of Science, 106, 10522

\bibitem[{{Brinch} {et~al.}(2009){Brinch}, {J{\o}rgensen}, \&
  {Hogerheijde}}]{brinch09}
{Brinch}, C., {J{\o}rgensen}, J.~K., \& {Hogerheijde}, M.~R. 2009, \aap, 502,
  199

\bibitem[{{Calcutt} {et~al.}(2018{\natexlab{a}}){Calcutt}, {J{\o}rgensen},
  {M{\"u}ller}, {Kristensen}, {Coutens}, {Bourke}, {Garrod}, {Persson}, {van
  der Wiel}, {van Dishoeck}, \& {Wampfler}}]{calcutt18a}
{Calcutt}, H., {J{\o}rgensen}, J.~K., {M{\"u}ller}, H.~S.~P., {et~al.}
  2018{\natexlab{a}}, ArXiv e-prints, arXiv:1804.09210

\bibitem[{{Calcutt} {et~al.}(2018{\natexlab{b}}){Calcutt}, {Fiechter},
  {Willis}, {M{\"u}ller}, {Garrod}, {J{\o}rgensen}, {Wampfler}, {Bourke},
  {Coutens}, {Drozdovskaya}, {Ligterink}, \& {Kristensen}}]{calcutt18b}
{Calcutt}, H., {Fiechter}, M.~R., {Willis}, E.~R., {et~al.} 2018{\natexlab{b}},
  ArXiv e-prints, arXiv:1807.02909

\bibitem[{{Caux} {et~al.}(2011){Caux}, {Kahane}, {Castets}, {Coutens},
  {Ceccarelli}, {Bacmann}, {Bisschop}, {Bottinelli}, {Comito}, {Helmich},
  {Lefloch}, {Parise}, {Schilke}, {Tielens}, {van Dishoeck}, {Vastel},
  {Wakelam}, \& {Walters}}]{caux11}
{Caux}, E., {Kahane}, C., {Castets}, A., {et~al.} 2011, \aap, 532, A23

\bibitem[{{Cazaux} {et~al.}(2003){Cazaux}, {Tielens}, {Ceccarelli}, {Castets},
  {Wakelam}, {Caux}, {Parise}, \& {Teyssier}}]{cazaux03}
{Cazaux}, S., {Tielens}, A.~G.~G.~M., {Ceccarelli}, C., {et~al.} 2003, \apjl,
  593, L51

\bibitem[{{Ceccarelli} {et~al.}(2010){Ceccarelli}, {Bacmann}, {Boogert},
  {Caux}, {Dominik}, {Lefloch}, {Lis}, {Schilke}, {van der Tak}, {Caselli},
  {Cernicharo}, {Codella}, {Comito}, {Fuente}, {Baudry}, {Bell}, {Benedettini},
  {Bergin}, {Blake}, {Bottinelli}, {Cabrit}, {Castets}, {Coutens}, {Crimier},
  {Demyk}, {Encrenaz}, {Falgarone}, {Gerin}, {Goldsmith}, {Helmich},
  {Hennebelle}, {Henning}, {Herbst}, {Hily-Blant}, {Jacq}, {Kahane}, {Kama},
  {Klotz}, {Langer}, {Lord}, {Lorenzani}, {Maret}, {Melnick}, {Neufeld},
  {Nisini}, {Pacheco}, {Pagani}, {Parise}, {Pearson}, {Phillips}, {Salez},
  {Saraceno}, {Schuster}, {Tielens}, {van der Wiel}, {Vastel}, {Viti},
  {Wakelam}, {Walters}, {Wyrowski}, {Yorke}, {Liseau}, {Olberg}, {Szczerba},
  {Benz}, \& {Melchior}}]{ceccarelli10}
{Ceccarelli}, C., {Bacmann}, A., {Boogert}, A., {et~al.} 2010, \aap, 521, L22

\bibitem[{{Cernicharo}(2012)}]{cernicharo12}
{Cernicharo}, J. 2012, in EAS Publications Series, Vol.~58, EAS Publications
  Series, ed. C.~{Stehl{\'e}}, C.~{Joblin}, \& L.~{d'Hendecourt}, 251--261

\bibitem[{{Cernicharo} {et~al.}(2016){Cernicharo}, {Kisiel}, {Tercero},
  {Kolesnikov{\'a}}, {Medvedev}, {L{\'o}pez}, {Fortman}, {Winnewisser}, {de
  Lucia}, {Alonso}, \& {Guillemin}}]{cernicharo16}
{Cernicharo}, J., {Kisiel}, Z., {Tercero}, B., {et~al.} 2016, \aap, 587, L4

\bibitem[{{Chandler} {et~al.}(2005){Chandler}, {Brogan}, {Shirley}, \&
  {Loinard}}]{chandler05}
{Chandler}, C.~J., {Brogan}, C.~L., {Shirley}, Y.~L., \& {Loinard}, L. 2005,
  \apj, 632, 371

\bibitem[{{Charnley} \& {Rodgers}(2008)}]{charnley08}
{Charnley}, S.~B., \& {Rodgers}, S.~D. 2008, \ssr, 138, 59

\bibitem[{{Chiar} {et~al.}(2013){Chiar}, {Tielens}, {Adamson}, \&
  {Ricca}}]{chiar13}
{Chiar}, J.~E., {Tielens}, A.~G.~G.~M., {Adamson}, A.~J., \& {Ricca}, A. 2013,
  \apj, 770, 78

\bibitem[{{Clayton} \& {Nittler}(2004)}]{clayton04}
{Clayton}, D.~D., \& {Nittler}, L.~R. 2004, Origin and Evolution of the
  Elements, 297

\bibitem[{{Colzi} {et~al.}(2018){Colzi}, {Fontani}, {Rivilla},
  {S{\'a}nchez-Monge}, {Testi}, {Beltr{\'a}n}, \& {Caselli}}]{colzi18}
{Colzi}, L., {Fontani}, F., {Rivilla}, V.~M., {et~al.} 2018, \mnras, 478, 3693

\bibitem[{{Coutens} {et~al.}(2012){Coutens}, {Vastel}, {Caux}, {Ceccarelli},
  {Bottinelli}, {Wiesenfeld}, {Faure}, {Scribano}, \& {Kahane}}]{coutens12}
{Coutens}, A., {Vastel}, C., {Caux}, E., {et~al.} 2012, \aap, 539, A132

\bibitem[{{Crimier} {et~al.}(2010){Crimier}, {Ceccarelli}, {Maret},
  {Bottinelli}, {Caux}, {Kahane}, {Lis}, \& {Olofsson}}]{crimier10}
{Crimier}, N., {Ceccarelli}, C., {Maret}, S., {et~al.} 2010, \aap, 519, A65

\bibitem[{{Crockett} {et~al.}(2014{\natexlab{a}}){Crockett}, {Bergin}, {Neill},
  {Black}, {Blake}, \& {Kleshcheva}}]{Crockett2014a}
{Crockett}, N.~R., {Bergin}, E.~A., {Neill}, J.~L., {et~al.}
  2014{\natexlab{a}}, \apj, 781, 114

\bibitem[{{Crockett} {et~al.}(2015){Crockett}, {Bergin}, {Neill}, {Favre},
  {Blake}, {Herbst}, {Anderson}, \& {Hassel}}]{Crockett15}
---. 2015, \apj, 806, 239

\bibitem[{{Crockett} {et~al.}(2010){Crockett}, {Bergin}, {Wang}, {Lis}, {Bell},
  {Blake}, {Boogert}, {Bumble}, {Cabrit}, {Caux}, {Ceccarelli}, {Cernicharo},
  {Comito}, {Daniel}, {Dubernet}, {Emprechtinger}, {Encrenaz}, {Falgarone},
  {Gerin}, {Giesen}, {Goicoechea}, {Goldsmith}, {Gupta}, {G{\"u}sten},
  {Hartogh}, {Helmich}, {Herbst}, {Honingh}, {Joblin}, {Johnstone}, {Karpov},
  {Kawamura}, {Kooi}, {Krieg}, {Langer}, {Latter}, {Lord}, {Maret}, {Martin},
  {Melnick}, {Menten}, {Morris}, {M{\"u}ller}, {Murphy}, {Neufeld},
  {Ossenkopf}, {Pearson}, {P{\'e}rault}, {Phillips}, {Plume}, {Qin},
  {Roelfsema}, {Schieder}, {Schilke}, {Schlemmer}, {Stutzki}, {van der Tak},
  {Tielens}, {Trappe}, {Vastel}, {Yorke}, {Yu}, \& {Zmuidzinas}}]{crockett10}
{Crockett}, N.~R., {Bergin}, E.~A., {Wang}, S., {et~al.} 2010, \aap, 521, L21

\bibitem[{{Crockett} {et~al.}(2014{\natexlab{b}}){Crockett}, {Bergin}, {Neill},
  {Favre}, {Schilke}, {Lis}, {Bell}, {Blake}, {Cernicharo}, {Emprechtinger},
  {Esplugues}, {Gupta}, {Kleshcheva}, {Lord}, {Marcelino}, {McGuire},
  {Pearson}, {Phillips}, {Plume}, {van der Tak}, {Tercero}, \&
  {Yu}}]{Crockett14b}
{Crockett}, N.~R., {Bergin}, E.~A., {Neill}, J.~L., {et~al.}
  2014{\natexlab{b}}, \apj, 787, 112

\bibitem[{{Dullemond} {et~al.}(2002){Dullemond}, {van Zadelhoff}, \&
  {Natta}}]{dullemond02}
{Dullemond}, C.~P., {van Zadelhoff}, G.~J., \& {Natta}, A. 2002, \aap, 389, 464

\bibitem[{{Dumouchel} {et~al.}(2010){Dumouchel}, {Faure}, \&
  {Lique}}]{dumouchel10}
{Dumouchel}, F., {Faure}, A., \& {Lique}, F. 2010, \mnras, 406, 2488

\bibitem[{{Dzib} {et~al.}(2018){Dzib}, {Ortiz-Le{\'o}n},
  {Hern{\'a}ndez-G{\'o}mez}, {Loinard}, {Mioduszewski}, {Claussen}, {Menten},
  {Caux}, \& {Sanna}}]{dzib18}
{Dzib}, S.~A., {Ortiz-Le{\'o}n}, G.~N., {Hern{\'a}ndez-G{\'o}mez}, A., {et~al.}
  2018, \aap, 614, A20

\bibitem[{{Endres} {et~al.}(2016){Endres}, {Schlemmer}, {Schilke}, {Stutzki},
  \& {M{\"u}ller}}]{endres16}
{Endres}, C.~P., {Schlemmer}, S., {Schilke}, P., {Stutzki}, J., \&
  {M{\"u}ller}, H.~S.~P. 2016, Journal of Molecular Spectroscopy, 327, 95

\bibitem[{Foreman-Mackey(2016)}]{Foreman-Mackey:2016}
Foreman-Mackey, D. 2016, corner.py: corner.py v1.0.2, , ,
  doi:10.5281/zenodo.45906

\bibitem[{{Foreman-Mackey} {et~al.}(2013{\natexlab{a}}){Foreman-Mackey},
  {Hogg}, {Lang}, \& {Goodman}}]{foreman-mackey13}
{Foreman-Mackey}, D., {Hogg}, D.~W., {Lang}, D., \& {Goodman}, J.
  2013{\natexlab{a}}, \pasp, 125, 306

\bibitem[{{Foreman-Mackey} {et~al.}(2013{\natexlab{b}}){Foreman-Mackey},
  {Conley}, {Meierjurgen Farr}, {Hogg}, {Lang}, {Marshall}, {Price-Whelan},
  {Sanders}, \& {Zuntz}}]{foreman-mackey13ascl}
{Foreman-Mackey}, D., {Conley}, A., {Meierjurgen Farr}, W., {et~al.}
  2013{\natexlab{b}}, {emcee: The MCMC Hammer}, Astrophysics Source Code
  Library, , , ascl:1303.002

\bibitem[{{Fray} {et~al.}(2016){Fray}, {Bardyn}, {Cottin}, {Altwegg},
  {Baklouti}, {Briois}, {Colangeli}, {Engrand}, {Fischer}, {Glasmachers},
  {Gr{\"u}n}, {Haerendel}, {Henkel}, {H{\"o}fner}, {Hornung}, {Jessberger},
  {Koch}, {Kr{\"u}ger}, {Langevin}, {Lehto}, {Lehto}, {Le Roy}, {Merouane},
  {Modica}, {Orthous-Daunay}, {Paquette}, {Raulin}, {Ryn{\"o}}, {Schulz},
  {Sil{\'e}n}, {Siljestr{\"o}m}, {Steiger}, {Stenzel}, {Stephan}, {Thirkell},
  {Thomas}, {Torkar}, {Varmuza}, {Wanczek}, {Zaprudin}, {Kissel}, \&
  {Hilchenbach}}]{fray16}
{Fray}, N., {Bardyn}, A., {Cottin}, H., {et~al.} 2016, \nat, 538, 72

\bibitem[{{Fray} {et~al.}(2017){Fray}, {Bardyn}, {Cottin}, {Baklouti},
  {Briois}, {Engrand}, {Fischer}, {Hornung}, {Isnard}, {Langevin}, {Lehto}, {Le
  Roy}, {Mellado}, {Merouane}, {Modica}, {Orthous-Daunay}, {Paquette},
  {Ryn{\"o}}, {Schulz}, {Sil{\'e}n}, {Siljestr{\"o}m}, {Stenzel}, {Thirkell},
  {Varmuza}, {Zaprudin}, {Kissel}, \& {Hilchenbach}}]{fray17}
---. 2017, \mnras, 469, S506

\bibitem[{{F{\"u}ri} \& {Marty}(2015)}]{furi15}
{F{\"u}ri}, E., \& {Marty}, B. 2015, Nature Geoscience, 8, 515

\bibitem[{{Furuya} \& {Aikawa}(2018)}]{furuya18}
{Furuya}, K., \& {Aikawa}, Y. 2018, \apj, 857, 105

\bibitem[{{Furuya} \& {Persson}(2018)}]{furuya18jun}
{Furuya}, K., \& {Persson}, M.~V. 2018, \mnras, 476, 4994

\bibitem[{{Furuya} {et~al.}(2018){Furuya}, {Watanabe}, {Sakai}, {Aikawa}, \&
  {Yamamoto}}]{furuya18b}
{Furuya}, K., {Watanabe}, Y., {Sakai}, T., {Aikawa}, Y., \& {Yamamoto}, S.
  2018, \aap, 615, L16

\bibitem[{{Garrod} \& {Herbst}(2006)}]{garrod06}
{Garrod}, R.~T., \& {Herbst}, E. 2006, \aap, 457, 927

\bibitem[{{Gildas Team}(2013)}]{gildas13}
{Gildas Team}. 2013, {GILDAS: Grenoble Image and Line Data Analysis Software},
  Astrophysics Source Code Library, , , ascl:1305.010

\bibitem[{Goesmann {et~al.}(2015)Goesmann, Rosenbauer, Bredeh{\"o}ft, Cabane,
  Ehrenfreund, Gautier, Giri, Kr{\"u}ger, Le~Roy, MacDermott, McKenna-Lawlor,
  Meierhenrich, Caro, Raulin, Roll, Steele, Steininger, Sternberg, Szopa,
  Thiemann, \& Ulamec}]{Goesmannaab0689}
Goesmann, F., Rosenbauer, H., Bredeh{\"o}ft, J.~H., {et~al.} 2015, Science,
  349, http://science.sciencemag.org/content/349/6247/aab0689.full.pdf

\bibitem[{{Gong} {et~al.}(2015){Gong}, {Henkel}, {Thorwirth}, {Spezzano},
  {Menten}, {Walmsley}, {Wyrowski}, {Mao}, \& {Klein}}]{gong15}
{Gong}, Y., {Henkel}, C., {Thorwirth}, S., {et~al.} 2015, \aap, 581, A48

\bibitem[{Goodman \& Weare(2010)}]{goodman10}
Goodman, J., \& Weare, J. 2010, Communications in applied mathematics and
  computational science, 5, 65

\bibitem[{{Guzm{\'a}n} {et~al.}(2017){Guzm{\'a}n}, {{\"O}berg}, {Huang},
  {Loomis}, \& {Qi}}]{guzman17}
{Guzm{\'a}n}, V.~V., {{\"O}berg}, K.~I., {Huang}, J., {Loomis}, R., \& {Qi}, C.
  2017, \apj, 836, 30

\bibitem[{{Hasegawa} \& {Herbst}(1993)}]{hasegawa93}
{Hasegawa}, T.~I., \& {Herbst}, E. 1993, \mnras, 261, 83

\bibitem[{{Hily-Blant} {et~al.}(2013){Hily-Blant}, {Bonal}, {Faure}, \&
  {Quirico}}]{hily-blant13}
{Hily-Blant}, P., {Bonal}, L., {Faure}, A., \& {Quirico}, E. 2013, \icarus,
  223, 582

\bibitem[{{Hily-Blant} {et~al.}(2017){Hily-Blant}, {Magalhaes}, {Kastner},
  {Faure}, {Forveille}, \& {Qi}}]{hily-blant17}
{Hily-Blant}, P., {Magalhaes}, V., {Kastner}, J., {et~al.} 2017, \aap, 603, L6

\bibitem[{{Hily-Blant} {et~al.}(2010){Hily-Blant}, {Maret}, {Bacmann},
  {Bottinelli}, {Parise}, {Caux}, {Faure}, {Bergin}, {Blake}, {Castets},
  {Ceccarelli}, {Cernicharo}, {Coutens}, {Crimier}, {Demyk}, {Dominik},
  {Gerin}, {Hennebelle}, {Henning}, {Kahane}, {Klotz}, {Melnick}, {Pagani},
  {Schilke}, {Vastel}, {Wakelam}, {Walters}, {Baudry}, {Bell}, {Benedettini},
  {Boogert}, {Cabrit}, {Caselli}, {Codella}, {Comito}, {Encrenaz}, {Falgarone},
  {Fuente}, {Goldsmith}, {Helmich}, {Herbst}, {Jacq}, {Kama}, {Langer},
  {Lefloch}, {Lis}, {Lord}, {Lorenzani}, {Neufeld}, {Nisini}, {Pacheco},
  {Phillips}, {Salez}, {Saraceno}, {Schuster}, {Tielens}, {van der Tak}, {van
  der Wiel}, {Viti}, {Wyrowski}, \& {Yorke}}]{hily-blant10}
{Hily-Blant}, P., {Maret}, S., {Bacmann}, A., {et~al.} 2010, \aap, 521, L52

\bibitem[{{Hogerheijde} \& {van der
  Tak}(2000{\natexlab{a}})}]{hogerheijde00ascl}
{Hogerheijde}, M., \& {van der Tak}, F. 2000{\natexlab{a}}, {RATRAN: Radiative
  Transfer and Molecular Excitation in One and Two Dimensions}, Astrophysics
  Source Code Library, , , ascl:0008.002

\bibitem[{{Hogerheijde} \& {van der Tak}(2000{\natexlab{b}})}]{hogerheijde00}
{Hogerheijde}, M.~R., \& {van der Tak}, F.~F.~S. 2000{\natexlab{b}}, \aap, 362,
  697

\bibitem[{{Hogg} \& {Foreman-Mackey}(2018)}]{hogg18}
{Hogg}, D.~W., \& {Foreman-Mackey}, D. 2018, \apjs, 236, 11

\bibitem[{{Hollenbach} {et~al.}(2009){Hollenbach}, {Kaufman}, {Bergin}, \&
  {Melnick}}]{hollenbach09}
{Hollenbach}, D., {Kaufman}, M.~J., {Bergin}, E.~A., \& {Melnick}, G.~J. 2009,
  \apj, 690, 1497

\bibitem[{Hunter(2007)}]{Hunter:2007}
Hunter, J.~D. 2007, Computing In Science \& Engineering, 9, 90

\bibitem[{{Jacobsen} {et~al.}(2017){Jacobsen}, {J{\o}rgensen}, {van der Wiel},
  {Calcutt}, {Bourke}, {Brinch}, {Coutens}, {Drozdovskaya}, {Kristensen},
  {M{\"u}ller}, \& {Wampfler}}]{jacobsen17}
{Jacobsen}, S.~K., {J{\o}rgensen}, J.~K., {van der Wiel}, M.~H.~D., {et~al.}
  2017, ArXiv e-prints, arXiv:1712.06984

\bibitem[{{Jensen} {et~al.}(2007){Jensen}, {Rachford}, \& {Snow}}]{jensen07}
{Jensen}, A.~G., {Rachford}, B.~L., \& {Snow}, T.~P. 2007, \apj, 654, 955

\bibitem[{{Jones}(2016)}]{jones16a}
{Jones}, A.~P. 2016, Royal Society Open Science, 3, 160221

\bibitem[{{J{\o}rgensen} {et~al.}(2011){J{\o}rgensen}, {Bourke}, {Nguyen
  Luong}, \& {Takakuwa}}]{jorgensen11}
{J{\o}rgensen}, J.~K., {Bourke}, T.~L., {Nguyen Luong}, Q., \& {Takakuwa}, S.
  2011, \aap, 534, A100

\bibitem[{{J{\o}rgensen} {et~al.}(2002){J{\o}rgensen}, {Sch{\"o}ier}, \& {van
  Dishoeck}}]{jorgensen02}
{J{\o}rgensen}, J.~K., {Sch{\"o}ier}, F.~L., \& {van Dishoeck}, E.~F. 2002,
  \aap, 389, 908

\bibitem[{{J{\o}rgensen} {et~al.}(2005){J{\o}rgensen}, {Sch{\"o}ier}, \& {van
  Dishoeck}}]{jorgensen05}
---. 2005, \aap, 435, 177

\bibitem[{{J{\o}rgensen} {et~al.}(2016){J{\o}rgensen}, {van der Wiel},
  {Coutens}, {Lykke}, {M{\"u}ller}, {van Dishoeck}, {Calcutt}, {Bjerkeli},
  {Bourke}, {Drozdovskaya}, {Favre}, {Fayolle}, {Garrod}, {Jacobsen},
  {{\"O}berg}, {Persson}, \& {Wampfler}}]{jorgensen16}
{J{\o}rgensen}, J.~K., {van der Wiel}, M.~H.~D., {Coutens}, A., {et~al.} 2016,
  \aap, 595, A117

\bibitem[{{Kahane} {et~al.}(2018){Kahane}, {Jaber Al-Edhari}, {Ceccarelli},
  {L{\'o}pez-Sepulcre}, {Fontani}, \& {Kama}}]{kahane18}
{Kahane}, C., {Jaber Al-Edhari}, A., {Ceccarelli}, C., {et~al.} 2018, \apj,
  852, 130

\bibitem[{{Kissel} \& {Krueger}(1987)}]{kissel87}
{Kissel}, J., \& {Krueger}, F.~R. 1987, \nat, 326, 755

\bibitem[{{Knauth} {et~al.}(2003){Knauth}, {Andersson}, {McCandliss}, \&
  {Moos}}]{knauth03}
{Knauth}, D.~C., {Andersson}, B.-G., {McCandliss}, S.~R., \& {Moos}, H.~W.
  2003, \apjl, 596, L51

\bibitem[{{Le Roy} {et~al.}(2015){Le Roy}, {Altwegg}, {Balsiger}, {Berthelier},
  {Bieler}, {Briois}, {Calmonte}, {Combi}, {De Keyser}, {Dhooghe}, {Fiethe},
  {Fuselier}, {Gasc}, {Gombosi}, {H{\"a}ssig}, {J{\"a}ckel}, {Rubin}, \&
  {Tzou}}]{leroy15}
{Le Roy}, L., {Altwegg}, K., {Balsiger}, H., {et~al.} 2015, \aap, 583, A1

\bibitem[{{Ligterink} {et~al.}(2017){Ligterink}, {Coutens}, {Kofman},
  {M{\"u}ller}, {Garrod}, {Calcutt}, {Wampfler}, {J{\o}rgensen}, {Linnartz}, \&
  {van Dishoeck}}]{ligterink17}
{Ligterink}, N.~F.~W., {Coutens}, A., {Kofman}, V., {et~al.} 2017, \mnras, 469,
  2219

\bibitem[{{Loinard} {et~al.}(2008){Loinard}, {Torres}, {Mioduszewski}, \&
  {Rodr{\'{\i}}guez}}]{loinard08}
{Loinard}, L., {Torres}, R.~M., {Mioduszewski}, A.~J., \& {Rodr{\'{\i}}guez},
  L.~F. 2008, \apjl, 675, L29

\bibitem[{{Looney} {et~al.}(2000){Looney}, {Mundy}, \& {Welch}}]{looney00}
{Looney}, L.~W., {Mundy}, L.~G., \& {Welch}, W.~J. 2000, \apj, 529, 477

\bibitem[{{Lykke} {et~al.}(2017){Lykke}, {Coutens}, {J{\o}rgensen}, {van der
  Wiel}, {Garrod}, {M{\"u}ller}, {Bjerkeli}, {Bourke}, {Calcutt},
  {Drozdovskaya}, {Favre}, {Fayolle}, {Jacobsen}, {{\"O}berg}, {Persson}, {van
  Dishoeck}, \& {Wampfler}}]{lykke17}
{Lykke}, J.~M., {Coutens}, A., {J{\o}rgensen}, J.~K., {et~al.} 2017, \aap, 597,
  A53

\bibitem[{{Maret} {et~al.}(2004){Maret}, {Ceccarelli}, {Caux}, {Tielens},
  {J{\o}rgensen}, {van Dishoeck}, {Bacmann}, {Castets}, {Lefloch}, {Loinard},
  {Parise}, \& {Sch{\"o}ier}}]{maret04}
{Maret}, S., {Ceccarelli}, C., {Caux}, E., {et~al.} 2004, \aap, 416, 577

\bibitem[{{Marty}(2012)}]{marty12}
{Marty}, B. 2012, Earth and Planetary Science Letters, 313, 56

\bibitem[{{Marty} {et~al.}(2010){Marty}, {Zimmermann}, {Burnard}, {Wieler},
  {Heber}, {Burnett}, {Wiens}, \& {Bochsler}}]{marty10}
{Marty}, B., {Zimmermann}, L., {Burnard}, P.~G., {et~al.} 2010, \gca, 74, 340

\bibitem[{{Marty} {et~al.}(2013){Marty}, {Zimmermann}, {Pujol}, {Burgess}, \&
  {Philippot}}]{marty13}
{Marty}, B., {Zimmermann}, L., {Pujol}, M., {Burgess}, R., \& {Philippot}, P.
  2013, Science, 342, 101

\bibitem[{{Marty} {et~al.}(2016){Marty}, {Avice}, {Sano}, {Altwegg},
  {Balsiger}, {H{\"a}ssig}, {Morbidelli}, {Mousis}, \& {Rubin}}]{marty16}
{Marty}, B., {Avice}, G., {Sano}, Y., {et~al.} 2016, Earth and Planetary
  Science Letters, 441, 91

\bibitem[{{Marty} {et~al.}(2017){Marty}, {Altwegg}, {Balsiger}, {Bar-Nun},
  {Bekaert}, {Berthelier}, {Bieler}, {Briois}, {Calmonte}, {Combi}, {De
  Keyser}, {Fiethe}, {Fuselier}, {Gasc}, {Gombosi}, {Hansen}, {H{\"a}ssig},
  {J{\"a}ckel}, {Kopp}, {Korth}, {Le Roy}, {Mall}, {Mousis}, {Owen},
  {R{\`e}me}, {Rubin}, {S{\'e}mon}, {Tzou}, {Waite}, \& {Wurz}}]{marty17}
{Marty}, B., {Altwegg}, K., {Balsiger}, H., {et~al.} 2017, Science, 356, 1069

\bibitem[{{McElroy} {et~al.}(2013){McElroy}, {Walsh}, {Markwick}, {Cordiner},
  {Smith}, \& {Millar}}]{mcelroy13}
{McElroy}, D., {Walsh}, C., {Markwick}, A.~J., {et~al.} 2013, \aap, 550, A36

\bibitem[{{McKeegan} {et~al.}(2006){McKeegan}, {Al{\'e}on}, {Bradley},
  {Brownlee}, {Busemann}, {Butterworth}, {Chaussidon}, {Fallon}, {Floss},
  {Gilmour}, {Gounelle}, {Graham}, {Guan}, {Heck}, {Hoppe}, {Hutcheon}, {Huth},
  {Ishii}, {Ito}, {Jacobsen}, {Kearsley}, {Leshin}, {Liu}, {Lyon}, {Marhas},
  {Marty}, {Matrajt}, {Meibom}, {Messenger}, {Mostefaoui}, {Mukhopadhyay},
  {Nakamura-Messenger}, {Nittler}, {Palma}, {Pepin}, {Papanastassiou},
  {Robert}, {Schlutter}, {Snead}, {Stadermann}, {Stroud}, {Tsou}, {Westphal},
  {Young}, {Ziegler}, {Zimmermann}, \& {Zinner}}]{mckeegan06}
{McKeegan}, K., {Al{\'e}on}, J., {Bradley}, J., {et~al.} 2006, Science, 314,
  1724

\bibitem[{{Messenger} {et~al.}(2005){Messenger}, {Keller}, \&
  {Lauretta}}]{messenger05}
{Messenger}, S., {Keller}, L.~P., \& {Lauretta}, D.~S. 2005, Science, 309, 737

\bibitem[{{Morbidelli} {et~al.}(2012){Morbidelli}, {Lunine}, {O'Brien},
  {Raymond}, \& {Walsh}}]{morbidelli12}
{Morbidelli}, A., {Lunine}, J.~I., {O'Brien}, D.~P., {Raymond}, S.~N., \&
  {Walsh}, K.~J. 2012, Annual Review of Earth and Planetary Sciences, 40, 251

\bibitem[{{M{\"u}ller} {et~al.}(2001){M{\"u}ller}, {Thorwirth}, {Roth}, \&
  {Winnewisser}}]{muller01}
{M{\"u}ller}, H.~S.~P., {Thorwirth}, S., {Roth}, D.~A., \& {Winnewisser}, G.
  2001, \aap, 370, L49

\bibitem[{{Mumma} \& {Charnley}(2011)}]{mumma11}
{Mumma}, M.~J., \& {Charnley}, S.~B. 2011, \araa, 49, 471

\bibitem[{{Neill} {et~al.}(2013{\natexlab{a}}){Neill}, {Crockett}, {Bergin},
  {Pearson}, \& {Xu}}]{neill13b}
{Neill}, J.~L., {Crockett}, N.~R., {Bergin}, E.~A., {Pearson}, J.~C., \& {Xu},
  L.-H. 2013{\natexlab{a}}, \apj, 777, 85

\bibitem[{{Neill} {et~al.}(2013{\natexlab{b}}){Neill}, {Wang}, {Bergin},
  {Crockett}, {Favre}, {Plume}, \& {Melnick}}]{neill13a}
{Neill}, J.~L., {Wang}, S., {Bergin}, E.~A., {et~al.} 2013{\natexlab{b}}, \apj,
  770, 142

\bibitem[{{Nieva} \& {Przybilla}(2012)}]{nieva12}
{Nieva}, M.-F., \& {Przybilla}, N. 2012, \aap, 539, A143

\bibitem[{{Noble} {et~al.}(2013){Noble}, {Theule}, {Borget}, {Danger},
  {Chomat}, {Duvernay}, {Mispelaer}, \& {Chiavassa}}]{noble13}
{Noble}, J.~A., {Theule}, P., {Borget}, F., {et~al.} 2013, \mnras, 428, 3262

\bibitem[{{Parise} {et~al.}(2005){Parise}, {Caux}, {Castets}, {Ceccarelli},
  {Loinard}, {Tielens}, {Bacmann}, {Cazaux}, {Comito}, {Helmich}, {Kahane},
  {Schilke}, {van Dishoeck}, {Wakelam}, \& {Walters}}]{parise05}
{Parise}, B., {Caux}, E., {Castets}, A., {et~al.} 2005, \aap, 431, 547

\bibitem[{{Persson} {et~al.}(2013){Persson}, {J{\o}rgensen}, \& {van
  Dishoeck}}]{persson13}
{Persson}, M.~V., {J{\o}rgensen}, J.~K., \& {van Dishoeck}, E.~F. 2013, \aap,
  549, L3

\bibitem[{{Pety}(2005)}]{pety05}
{Pety}, J. 2005, in SF2A-2005: Semaine de l'Astrophysique Francaise, ed.
  F.~{Casoli}, T.~{Contini}, J.~M. {Hameury}, \& L.~{Pagani}, 721

\bibitem[{{Pety}(2018)}]{pety18}
{Pety}, J. 2018, in Submillimetre Single-dish Data Reduction and Array
  Combination Techniques, 11

\bibitem[{{Plume} {et~al.}(2012){Plume}, {Bergin}, {Phillips}, {Lis}, {Wang},
  {Crockett}, {Caux}, {Comito}, {Goldsmith}, \& {Schilke}}]{plume12}
{Plume}, R., {Bergin}, E.~A., {Phillips}, T.~G., {et~al.} 2012, \apj, 744, 28

\bibitem[{{Polanyi} \& {Wigner}(1925)}]{polanyi25}
{Polanyi}, M., \& {Wigner}, E. 1925, Zeitschrift fur Physik, 33, 429

\bibitem[{{Rodgers} \& {Charnley}(2003)}]{rodgers03}
{Rodgers}, S.~D., \& {Charnley}, S.~B. 2003, \apj, 585, 355

\bibitem[{{Roelfsema} {et~al.}(2012){Roelfsema}, {Helmich}, {Teyssier},
  {Ossenkopf}, {Morris}, {Olberg}, {Shipman}, {Risacher}, {Akyilmaz},
  {Assendorp}, {Avruch}, {Beintema}, {Biver}, {Boogert}, {Borys}, {Braine},
  {Caris}, {Caux}, {Cernicharo}, {Coeur-Joly}, {Comito}, {de Lange},
  {Delforge}, {Dieleman}, {Dubbeldam}, {de Graauw}, {Edwards}, {Fich},
  {Flederus}, {Gal}, {di Giorgio}, {Herpin}, {Higgins}, {Hoac}, {Huisman},
  {Jarchow}, {Jellema}, {de Jonge}, {Kester}, {Klein}, {Kooi}, {Kramer},
  {Laauwen}, {Larsson}, {Leinz}, {Lord}, {Lorenzani}, {Luinge}, {Marston},
  {Mart{\'{\i}}n-Pintado}, {McCoey}, {Melchior}, {Michalska}, {Moreno},
  {M{\"u}ller}, {Nowosielski}, {Okada}, {Orlea{\'n}ski}, {Phillips}, {Pearson},
  {Rabois}, {Ravera}, {Rector}, {Rengel}, {Sagawa}, {Salomons},
  {S{\'a}nchez-Su{\'a}rez}, {Schieder}, {Schl{\"o}der}, {Schm{\"u}lling},
  {Soldati}, {Stutzki}, {Thomas}, {Tielens}, {Vastel}, {Wildeman}, {Xie},
  {Xilouris}, {Wafelbakker}, {Whyborn}, {Zaal}, {Bell}, {Bjerkeli}, {De Beck},
  {Cavali{\'e}}, {Crockett}, {Hily-Blant}, {Kama}, {Kaminski}, {Lefl{\'o}ch},
  {Lombaert}, {de Luca}, {Makai}, {Marseille}, {Nagy}, {Pacheco}, {van der
  Wiel}, {Wang}, \& {Y{\i}ld{\i}z}}]{roelfsema12}
{Roelfsema}, P.~R., {Helmich}, F.~P., {Teyssier}, D., {et~al.} 2012, \aap, 537,
  A17

\bibitem[{Rohatgi \& ZlatanStanojevic(2017)}]{ankit_rohatgi_2017_802310}
Rohatgi, A., \& ZlatanStanojevic. 2017, ankitrohatgi/WebPlotDigitizer: Version
  3.12, , , doi:10.5281/zenodo.802310

\bibitem[{{Romano} {et~al.}(2017){Romano}, {Matteucci}, {Zhang},
  {Papadopoulos}, \& {Ivison}}]{romano17}
{Romano}, D., {Matteucci}, F., {Zhang}, Z.-Y., {Papadopoulos}, P.~P., \&
  {Ivison}, R.~J. 2017, \mnras, 470, 401

\bibitem[{{Roueff} {et~al.}(2015){Roueff}, {Loison}, \& {Hickson}}]{roueff15}
{Roueff}, E., {Loison}, J.~C., \& {Hickson}, K.~M. 2015, \aap, 576, A99

\bibitem[{{Rubin} {et~al.}(2015){Rubin}, {Altwegg}, {Balsiger}, {Bar-Nun},
  {Berthelier}, {Bieler}, {Bochsler}, {Briois}, {Calmonte}, {Combi}, {De
  Keyser}, {Dhooghe}, {Eberhardt}, {Fiethe}, {Fuselier}, {Gasc}, {Gombosi},
  {Hansen}, {H{\"a}ssig}, {J{\"a}ckel}, {Kopp}, {Korth}, {Le Roy}, {Mall},
  {Marty}, {Mousis}, {Owen}, {R{\`e}me}, {S{\'e}mon}, {Tzou}, {Waite}, \&
  {Wurz}}]{rubin15}
{Rubin}, M., {Altwegg}, K., {Balsiger}, H., {et~al.} 2015, Science, 348, 232

\bibitem[{{Sandford} {et~al.}(2006){Sandford}, {Al{\'e}on}, {Alexander},
  {Araki}, {Bajt}, {Baratta}, {Borg}, {Bradley}, {Brownlee}, {Brucato},
  {Burchell}, {Busemann}, {Butterworth}, {Clemett}, {Cody}, {Colangeli},
  {Cooper}, {D'Hendecourt}, {Djouadi}, {Dworkin}, {Ferrini}, {Fleckenstein},
  {Flynn}, {Franchi}, {Fries}, {Gilles}, {Glavin}, {Gounelle}, {Grossemy},
  {Jacobsen}, {Keller}, {Kilcoyne}, {Leitner}, {Matrajt}, {Meibom}, {Mennella},
  {Mostefaoui}, {Nittler}, {Palumbo}, {Papanastassiou}, {Robert}, {Rotundi},
  {Snead}, {Spencer}, {Stadermann}, {Steele}, {Stephan}, {Tsou}, {Tyliszczak},
  {Westphal}, {Wirick}, {Wopenka}, {Yabuta}, {Zare}, \&
  {Zolensky}}]{sandford06}
{Sandford}, S.~A., {Al{\'e}on}, J., {Alexander}, C.~M.~O.~., {et~al.} 2006,
  Science, 314, 1720

\bibitem[{{Sault} {et~al.}(1995){Sault}, {Teuben}, \& {Wright}}]{sault95}
{Sault}, R.~J., {Teuben}, P.~J., \& {Wright}, M.~C.~H. 1995, in Astronomical
  Society of the Pacific Conference Series, Vol.~77, Astronomical Data Analysis
  Software and Systems IV, ed. R.~A. {Shaw}, H.~E. {Payne}, \& J.~J.~E.
  {Hayes}, 433

\bibitem[{{Sch{\"o}ier} {et~al.}(2002){Sch{\"o}ier}, {J{\o}rgensen}, {van
  Dishoeck}, \& {Blake}}]{schoier02}
{Sch{\"o}ier}, F.~L., {J{\o}rgensen}, J.~K., {van Dishoeck}, E.~F., \& {Blake},
  G.~A. 2002, \aap, 390, 1001

\bibitem[{{Sch{\"o}ier} {et~al.}(2005){Sch{\"o}ier}, {van der Tak}, {van
  Dishoeck}, \& {Black}}]{schoier05}
{Sch{\"o}ier}, F.~L., {van der Tak}, F.~F.~S., {van Dishoeck}, E.~F., \&
  {Black}, J.~H. 2005, \aap, 432, 369

\bibitem[{{Schwarz} \& {Bergin}(2014)}]{schwarz14}
{Schwarz}, K.~R., \& {Bergin}, E.~A. 2014, \apj, 797, 113

\bibitem[{{Shu}(1977)}]{shu77}
{Shu}, F.~H. 1977, \apj, 214, 488

\bibitem[{Sz\H{o}ri \& Jedlovszky(2014)}]{szőri14}
Sz\H{o}ri, M., \& Jedlovszky, P. 2014, The Journal of Physical Chemistry C,
  118, 3599

\bibitem[{{Takakuwa} {et~al.}(2007){Takakuwa}, {Ohashi}, {Bourke}, {Hirano},
  {Ho}, {J{\o}rgensen}, {Kuan}, {Wilner}, \& {Yeh}}]{takakuwa07}
{Takakuwa}, S., {Ohashi}, N., {Bourke}, T.~L., {et~al.} 2007, \apj, 662, 431

\bibitem[{{The Astropy Collaboration} {et~al.}(2018){The Astropy
  Collaboration}, {Price-Whelan}, {Sip{\H o}cz}, {G{\"u}nther}, {Lim},
  {Crawford}, {Conseil}, {Shupe}, {Craig}, {Dencheva}, {Ginsburg},
  {VanderPlas}, {Bradley}, {P{\'e}rez-Su{\'a}rez}, {de Val-Borro}, {Aldcroft},
  {Cruz}, {Robitaille}, {Tollerud}, {Ardelean}, {Babej}, {Bachetti}, {Bakanov},
  {Bamford}, {Barentsen}, {Barmby}, {Baumbach}, {Berry}, {Biscani}, {Boquien},
  {Bostroem}, {Bouma}, {Brammer}, {Bray}, {Breytenbach}, {Buddelmeijer},
  {Burke}, {Calderone}, {Cano Rodr{\'{\i}}guez}, {Cara}, {Cardoso},
  {Cheedella}, {Copin}, {Crichton}, {D{\'A}vella}, {Deil}, {Depagne},
  {Dietrich}, {Donath}, {Droettboom}, {Earl}, {Erben}, {Fabbro}, {Ferreira},
  {Finethy}, {Fox}, {Garrison}, {Gibbons}, {Goldstein}, {Gommers}, {Greco},
  {Greenfield}, {Groener}, {Grollier}, {Hagen}, {Hirst}, {Homeier}, {Horton},
  {Hosseinzadeh}, {Hu}, {Hunkeler}, {Ivezi{\'c}}, {Jain}, {Jenness}, {Kanarek},
  {Kendrew}, {Kern}, {Kerzendorf}, {Khvalko}, {King}, {Kirkby}, {Kulkarni},
  {Kumar}, {Lee}, {Lenz}, {Littlefair}, {Ma}, {Macleod}, {Mastropietro},
  {McCully}, {Montagnac}, {Morris}, {Mueller}, {Mumford}, {Muna}, {Murphy},
  {Nelson}, {Nguyen}, {Ninan}, {N{\"o}the}, {Ogaz}, {Oh}, {Parejko}, {Parley},
  {Pascual}, {Patil}, {Patil}, {Plunkett}, {Prochaska}, {Rastogi}, {Reddy
  Janga}, {Sabater}, {Sakurikar}, {Seifert}, {Sherbert}, {Sherwood-Taylor},
  {Shih}, {Sick}, {Silbiger}, {Singanamalla}, {Singer}, {Sladen}, {Sooley},
  {Sornarajah}, {Streicher}, {Teuben}, {Thomas}, {Tremblay}, {Turner},
  {Terr{\'o}n}, {van Kerkwijk}, {de la Vega}, {Watkins}, {Weaver}, {Whitmore},
  {Woillez}, \& {Zabalza}}]{astropy18}
{The Astropy Collaboration}, {Price-Whelan}, A.~M., {Sip{\H o}cz}, B.~M.,
  {et~al.} 2018, ArXiv e-prints, arXiv:1801.02634

\bibitem[{{Tielens} \& {Charnley}(2013)}]{tielens13}
{Tielens}, A., \& {Charnley}, S. 2013, in Planetary and Interstellar Processes
  Relevant to the Origins of Life, ed. D.~Whittet (Springer Netherlands),
  25--51

\bibitem[{{van der Tak} \& {Hogerheijde}(2007)}]{vandertak07a}
{van der Tak}, F., \& {Hogerheijde}, M. 2007, ArXiv Astrophysics e-prints,
  astro-ph/0702385

\bibitem[{{van Dishoeck} {et~al.}(1993){van Dishoeck}, {Blake}, {Draine}, \&
  {Lunine}}]{vandishoeck93}
{van Dishoeck}, E.~F., {Blake}, G.~A., {Draine}, B.~T., \& {Lunine}, J.~I.
  1993, in Protostars and Planets III, ed. E.~H. {Levy} \& J.~I. {Lunine},
  163--241

\bibitem[{{van Dishoeck} {et~al.}(1995){van Dishoeck}, {Blake}, {Jansen}, \&
  {Groesbeck}}]{vandishoeck95}
{van Dishoeck}, E.~F., {Blake}, G.~A., {Jansen}, D.~J., \& {Groesbeck}, T.~D.
  1995, \apj, 447, 760

\bibitem[{{van Kooten} {et~al.}(2017){van Kooten}, {Nagashima}, {Kasama},
  {Wampfler}, {Ramsey}, {Frimann}, {Balogh}, {Schiller}, {Wielandt}, {Franchi},
  {J{\o}rgensen}, {Krot}, \& {Bizzarro}}]{vankooten17}
{van Kooten}, E.~M.~M.~E., {Nagashima}, K., {Kasama}, T., {et~al.} 2017, \gca,
  205, 119

\bibitem[{{Visser} {et~al.}(2013){Visser}, {J{\o}rgensen}, {Kristensen}, {van
  Dishoeck}, \& {Bergin}}]{visser13}
{Visser}, R., {J{\o}rgensen}, J.~K., {Kristensen}, L.~E., {van Dishoeck},
  E.~F., \& {Bergin}, E.~A. 2013, \apj, 769, 19

\bibitem[{{Wakelam} {et~al.}(2017){Wakelam}, {Loison}, {Mereau}, \&
  {Ruaud}}]{wakelam17}
{Wakelam}, V., {Loison}, J.-C., {Mereau}, R., \& {Ruaud}, M. 2017, Molecular
  Astrophysics, 6, 22

\bibitem[{{Wakelam} {et~al.}(2012){Wakelam}, {Herbst}, {Loison}, {Smith},
  {Chandrasekaran}, {Pavone}, {Adams}, {Bacchus-Montabonel}, {Bergeat},
  {B{\'e}roff}, {Bierbaum}, {Chabot}, {Dalgarno}, {van Dishoeck}, {Faure},
  {Geppert}, {Gerlich}, {Galli}, {H{\'e}brard}, {Hersant}, {Hickson},
  {Honvault}, {Klippenstein}, {Le Picard}, {Nyman}, {Pernot}, {Schlemmer},
  {Selsis}, {Sims}, {Talbi}, {Tennyson}, {Troe}, {Wester}, \&
  {Wiesenfeld}}]{wakelam12}
{Wakelam}, V., {Herbst}, E., {Loison}, J.-C., {et~al.} 2012, \apjs, 199, 21

\bibitem[{{Wampfler} {et~al.}(2014){Wampfler}, {J{\o}rgensen}, {Bizzarro}, \&
  {Bisschop}}]{wampfler14}
{Wampfler}, S.~F., {J{\o}rgensen}, J.~K., {Bizzarro}, M., \& {Bisschop}, S.~E.
  2014, \aap, 572, A24

\bibitem[{{Wang} {et~al.}(2011){Wang}, {Bergin}, {Crockett}, {Goldsmith},
  {Lis}, {Pearson}, {Schilke}, {Bell}, {Comito}, {Blake}, {Caux}, {Ceccarelli},
  {Cernicharo}, {Daniel}, {Dubernet}, {Emprechtinger}, {Encrenaz}, {Gerin},
  {Giesen}, {Goicoechea}, {Gupta}, {Herbst}, {Joblin}, {Johnstone}, {Langer},
  {Latter}, {Lord}, {Maret}, {Martin}, {Melnick}, {Menten}, {Morris},
  {M{\"u}ller}, {Murphy}, {Neufeld}, {Ossenkopf}, {P{\'e}rault}, {Phillips},
  {Plume}, {Qin}, {Schlemmer}, {Stutzki}, {Trappe}, {van der Tak}, {Vastel},
  {Yorke}, {Yu}, \& {Zmuidzinas}}]{wang11}
{Wang}, S., {Bergin}, E.~A., {Crockett}, N.~R., {et~al.} 2011, \aap, 527, A95

\bibitem[{Whittet(2013)}]{whittet2013planetary}
Whittet, D. 2013, Planetary and Interstellar Processes Relevant to the Origins
  of Life (Springer Netherlands)

\bibitem[{{Whittet}(2010)}]{whittet10}
{Whittet}, D.~C.~B. 2010, \apj, 710, 1009

\bibitem[{{Whittet} {et~al.}(2013){Whittet}, {Poteet}, {Chiar}, {Pagani},
  {Bajaj}, {Horne}, {Shenoy}, \& {Adamson}}]{whittet13}
{Whittet}, D.~C.~B., {Poteet}, C.~A., {Chiar}, J.~E., {et~al.} 2013, \apj, 774,
  102

\bibitem[{{Wilson}(1999)}]{wilson99}
{Wilson}, T.~L. 1999, Reports on Progress in Physics, 62, 143

\bibitem[{{Wirstr{\"o}m} \& {Charnley}(2018)}]{wirstrom18}
{Wirstr{\"o}m}, E.~S., \& {Charnley}, S.~B. 2018, \mnras, 474, 3720

\bibitem[{{Wirstr{\"o}m} {et~al.}(2012){Wirstr{\"o}m}, {Charnley}, {Cordiner},
  \& {Milam}}]{wirstrom12}
{Wirstr{\"o}m}, E.~S., {Charnley}, S.~B., {Cordiner}, M.~A., \& {Milam}, S.~N.
  2012, \apjl, 757, L11

\bibitem[{{Wright} {et~al.}(2015){Wright}, {Sheridan}, {Barber}, {Morgan},
  {Andrews}, \& {Morse}}]{wright15}
{Wright}, I.~P., {Sheridan}, S., {Barber}, S.~J., {et~al.} 2015, Science, 349,
  doi:10.1126/science.aab0673

\bibitem[{{Wyckoff} {et~al.}(1991){Wyckoff}, {Tegler}, \& {Engel}}]{wyckoff91}
{Wyckoff}, S., {Tegler}, S.~C., \& {Engel}, L. 1991, \apj, 367, 641

\bibitem[{{Zapata} {et~al.}(2013){Zapata}, {Loinard}, {Rodr{\'{\i}}guez},
  {Hern{\'a}ndez-Hern{\'a}ndez}, {Takahashi}, {Trejo}, \& {Parise}}]{zapata13}
{Zapata}, L.~A., {Loinard}, L., {Rodr{\'{\i}}guez}, L.~F., {et~al.} 2013,
  \apjl, 764, L14

\bibitem[{{Zernickel} {et~al.}(2012){Zernickel}, {Schilke}, {Schmiedeke},
  {Lis}, {Brogan}, {Ceccarelli}, {Comito}, {Emprechtinger}, {Hunter}, \&
  {M{\"o}ller}}]{zernickel12}
{Zernickel}, A., {Schilke}, P., {Schmiedeke}, A., {et~al.} 2012, \aap, 546, A87

\end{thebibliography}

\end{document}